\documentclass[fleqn,usenatbib]{mnras}
\usepackage{newtxtext,newtxmath}
\usepackage[T1]{fontenc}
\DeclareRobustCommand{\VAN}[3]{#2}
\let\VANthebibliography\thebibliography
\def\thebibliography{\DeclareRobustCommand{\VAN}[3]{##3}\VANthebibliography}
\usepackage{graphicx}
\usepackage{amsmath}
\usepackage{pdflscape}

\title[LAMOST J045019.27+394758.7:   possibly a Sculptor dwarf galaxy escapee \thanks]{LAMOST J045019.27+394758.7,  with peculiar abundances of N,  Na, V, Zn, is possibly a Sculptor dwarf galaxy escapee \thanks{Based on data collected using HCT/HESP}}

\author[Meenakshi Purandardas  et al.]{Meenakshi Purandardas$^{1,2}$, Aruna Goswami$^{1,3}$ \thanks{E-mail: aruna@iiap.res.in}, J. Shejeelammal$^{1}$,  Mayani Sonamben$^{1,3}$, 
\newauthor Ganesh Pawar$^{4}$, David Mkrtichian$^{5}$, Vijayakumar H. Doddamani$^{2}$, Santosh Joshi$^{4}$ \\
    $^{1}$Indian Institute of Astrophysics, Koramangala, Bangalore 560034,
    India;\\ 
$^{2}$ Department of physics, Bangalore university, Jnana Bharathi Campus, Karntaka 560056, India\\
$^{3}$ Institute of Frontier Science and Application, Bangalore, Karnataka, India\\
$^{4}$ Aryabhatta Research Institute of Observational Sciences (ARIES), Manora Peak, Nainital-263001, Uttarakhand, India.  \\
$^{5}$ National Astronomical Research Institute of Thailand (NARIT) 260 Moo 4, T. Donkaew, A. Maerim, Chiangmai, 50180 Thailand 
}

\date{Accepted 2022 April 22. Received 2022 April 22; in original form 2021 September 14}

\pubyear{2021}

\begin{document}
\label{firstpage}
\label{lastpage}
\pagerange{\pageref{firstpage}--\pageref{lastpage}}
\maketitle

\begin{abstract}
We present the results of the high-resolution (R$\sim$60,000) spectroscopic analysis of the star LAMOSTJ045019.27+394758.7 (hereafter J045) from the list of carbon stars of LAMOST DR2. From our analysis, we find that J045 does not exhibit the spectral characteristics of carbon stars. It is found to be a
metal-poor ( [Fe/H] = $-$1.05) giant that shows very unusual elemental abundances, particularly for
 N, Na, V, and  Zn.  J045 shows ${\alpha}$-elements (Mg, Si, Ca) with near-solar values ($<$[$\alpha$/Fe]$>$ = 0.09) in contrast to Galactic stars  that show [$\alpha$/Fe] in the range 0.2 to 0.3 dex.  In J045,  Sc and Ti are under abundant with [X/Fe] $\le$ $-$0.25. 
 Vanadium gives [V/Fe] = 0.51 and zinc is under-abundant with [Zn/Fe] = $-$0.62. The object exhibits near-solar abundances for Sr, Y, Ba, Pr, and Sm.  The La is marginally enhanced, and  Ce and Nd are marginally under-abundant in J045.  With  [Ba/Eu] = $-$0.38,  the object falls into the category of neutron-capture rich r-I stars. The estimated abundances of various elements  show that  the
 observed abundance pattern is not compatible with the abundances characteristic of Galactic metal-poor stars  but  matches quite closely  with the abundance pattern  of Sculptor Dwarf galaxy stars of similar metallicity. Based on the above observational evidences,  we suggest  that the object  is a possible Sculptor Dwarf Galaxy escapee. 
\end{abstract}


\section{Introduction}
Carbon stars are a special class of objects  characterised by strong molecular  bands of carbon  such as  C$_{2}$, CH and CN in their spectra. Many surveys were conducted to explore carbon stars such as the First Byurakan Spectral Sky Survey \citep{Gigoyan_1998}, infrared objective-prism surveys (\citealt{Alksnis_2001},   Automatic Plate Measuring survey \citep{Totten_1998, Ibata_2001},  and references therein), and the Large Sky Area Multi-Object Fibre Spectroscopic Telescope (LAMOST, \citealt{Wu_2011, Bai_2016, Ji_2016}). Many studies have shown that  a large fraction of iron-deficient stars exhibit  enhancement of  carbon as well as neutron-capture elements and are the ideal candidates to  study the  mechanism(s) that produce  these elements. 
The chemical peculiarity of these objects promoted many spectroscopic studies of this group.
One of our primary goals is therefore  to explore carbon enhanced stars with signatures of enhanced  abundance of heavy elements from large  surveys for conducting such studies. 
\par Following the low-resolution spectroscopic analysis, \cite{Goswami_2005, Goswami_2007}; \cite{goswami_2010b}, identified a substantial fraction ($\sim$ 30\%) of CH stars from the list of  faint high-latitude carbon stars of \cite{Christlieb_2001a}.  Follow-up detailed chemical analysis  of some of these objects have shown  many of them to be carbon-enhanced metal-poor (CEMP)   stars in a metallicity 
range [Fe/H] $<$ $-$1,  comprising  of  CEMP-r/s, CEMP-s and CEMP-no stars \citep{Goswami_2006, Purandardas_2019a, Goswami_2021, Purandardas_2021b}.    
    \cite{Ji_2016} could  identify 894 carbon stars from LAMOST DR2, from the  measurement of the line indices of carbon molecular lines  using spectra with R $\sim$ 1800, and S/N $>$ 10, and classified them into various spectral sub-classes such as C-N, C-R, and C-H. We  performed a detailed chemical analysis for a few objects from this list and  showed that the estimated abundances of the
    CEMP-r/s star LAMOSTJ151003.74+305407.3 (hereafter J151) could be well explained by the model yields ([X/Fe]) of i-process nucleosynthesis  of heavy elements, and LAMOSTJ091608.81+230734.6 (hereafter J091) was identified as a  CH giant \citep{Shejeelammal_2021a}.
    A parametric model based analysis have also shown that  the model yields ([X/Fe]) of i-process
    nucleosynthesis of heavy elements could well explain the estimated  abundance patterns of J151.  The 
    estimated  neutron-density-dependent [Rb/Zr] ratio confirmed former low-mass  asymptotic giant branch companions  for these two objects. Our kinematic analysis have  shown that  J151 is a halo object and  J091 belongs to the disk population \citep{Shejeelammal_2021a}.
 
 We report here the results based on the high-resolution abundance analysis  of another object  J045 from  \cite{Ji_2016} which  is  classified as a carbon star with no identifiable sub-type. 
Xiang et al. (2019) derived the atmospheric parameters and elemental abundances for O, C, N, Mg, Na, Si, Ti, Ca, Cr, Co, Mn, Ni, and Ba based on low-resolution (R$\sim$ 1800) spectroscopic analysis and using data driven models. Based on near-infrared (15140-16940 \AA\, ) spectroscopy using a spectra  at a  resolution
of R ${\sim}$ 22,500, Jonsson et al. (2020) determined the atmospheric parameters as well as these elemental  abundances  except for  Ba using APOGEE Stellar Parameter and Chemical Abundance pipeline (ASPCAP). 
 While \cite{Xiang_2019} estimated the atmospheric parameters T$\rm_{eff}$, log {g}, microturbulent velocity ${\zeta}$, and metallicity  [Fe/H] 
for this object as (4627 K, 2.26, 0.72 km s$^{-1}$,  $-$0.90),  
these estimates derived by \cite{Jonsson_2020} are respectively (4250 K, 1.00, 1.55 km s$^{-1}$ ,  $-$0.50). The large difference in the estimates of the atmospheric parameters by these two groups 
have prompted us to re-visit this object.  Although some of the aspects  were discussed in these two studies, in this work, the results from a high-resolution spectroscopic analysis  (R ${\sim}$ 60 000) of this object is presented   for the first time. 
In addition to the elements for which the abundances are available in these two studies, we have determined the abundances for  a few more elements  such as Eu, Sm, Nd, Pr, La, Sr, Y, and Zn. 
 Our analysis, however,  shows that this object does not exhibit the spectral properties of  carbon stars. From a visual inspection of its spectrum,  the  molecular bands due to carbon are  found to be very weak, and also  enhanced signature of neutron-capture elements were absent in the spectrum.   We find 
 a metallicity [Fe/H] = $-$1.05 for this object with elemental abundances that are quite unusual from that expected for Galactic metal-poor stars of similar metallicity.   The observed abundance pattern is however found to be   compatible  with the abundance patterns  observed in Sculptor Dwarf galaxy stars, indicating  that the object  J045 is possibly a Sculptor Dwarf Galaxy escapee. 
 We have also examined for signatures  of any internal mixing in this object using $^{12}$C/$^{13}$C and  [C/N]  ratios that were not discussed in  the previous works.
 
\par The paper is arranged as follows: Section \ref{section data} discusses observation and data reduction.
 Estimation  of stellar atmospheric parameters, mass, and age determination are discussed in Section \ref{section atmo para}. A discussion on the determination of abundances is presented in Section \ref{section abundance deter}.  Uncertainty in the abundance estimates is discussed in Section \ref{section abundance uncertainty}. Results of the TESS photometry for the programme star is presented in Section \ref{section tess}. Results obtained from the kinematic analysis are discussed  in Section \ref{section kinematic}. A discussion on the abundance analysis is given  in Section \ref{section discussion} and the conclusions are  presented in Section \ref{section conclusion}.

\section{Observations and Data Reduction} \label{section data}
The object J045  is taken  from the sample of carbon stars listed in   LAMOST DR2 \citep{Ji_2016}. The high-resolution (R $\sim$ 60,000) spectrum of the object is obtained from the Indian Astronomical Observatory (IAO), Hanle, using the HESP (high-resolution fiber-fed Hanle Echelle Spectrograph (HESP)) attached to 2m HCT (Himalayan Chandra Telescope). We have acquired two frames with exposure 2700s each.  The spectra are combined to get the final spectrum with high S/N ratio. The spectrum covers 
from 3530 to 9970 {\rm \AA} in the wavelength range. 

\begin{figure}
\includegraphics[width=9cm,height=9cm]{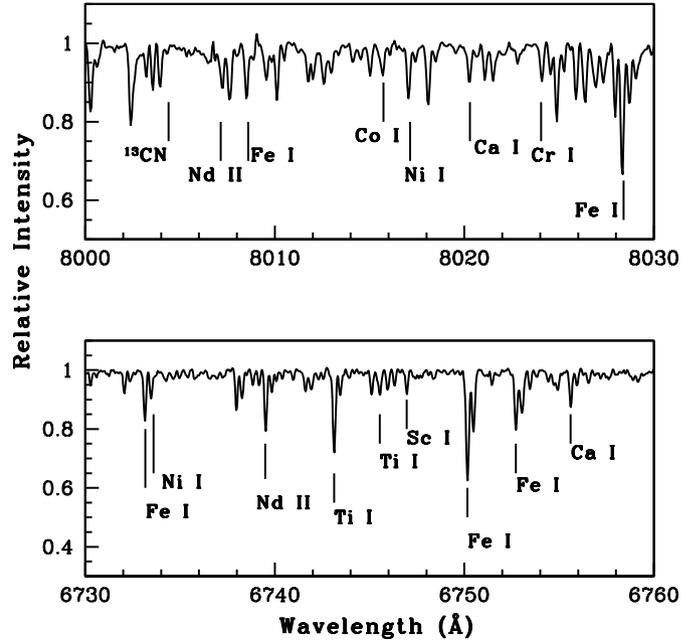} 
\caption{ Examples of  sample spectra of J045.  The wavelength region 8000 to 8030 \AA\, 
 is shown in the Upper panel and  the region 6730 to 6760 \AA\, is shown in the Lower panel.} \label{sample spectra}
\end{figure}

The spectrograph permits a resolution of 30,000 without slicer, and a resolution of 60,000 with slicer. 
The spectrum is recorded on a CCD with 4096$\times$4096 pixels,  each pixel  of 15 micron size. The data reduction is performed following the standard procedures using spectroscopic data reduction packages of  IRAF\footnote{IRAF is distributed by the National Optical Astronomical Observatories, which is operated by the Association for Universities for Research in Astronomy, Inc., under contract to the National Science Foundation}. Basic data of the programme star is presented in Table \ref{basic data}. 
Examples of sample spectra of J045 are  shown in Figure \ref{sample spectra}.

{\footnotesize
\begin{table*}
\caption{\bf LAMOST J045019.27+394758.7: Basic data} \label{basic data}
\resizebox{1\textwidth}{!}{\begin{tabular}{lccccccccccc}
\hline
RA$(2000)$   &Dec.$(2000)$  & Parallax &  G$^{a}$  & J$^{b}$  &H$^{b}$  &K$^{b}$ &Exposure    & S/N    &    & Date of obs. \\
            &               &   (mas)  &            &         &         &        & (seconds)  &   (at 4750 \AA) & (at 6000 \AA)     &  \\
\hline
 04 50 19.27 & 39 47 58.74 &   0.08$\pm$0.02    & 14.63   &  12.19   & 11.46    & 11.26 & 2700(2) & 57 & 72 & 29-09-2018 \\
\hline
\end{tabular}}

\textit{Note}. The number in the parenthesis with exposures indicates the number of frames taken, 
$^{a}$ Taken from SIMBAD Gaia DR2, $^{b}$ 2MASS, \cite{Cutri_2003}, 
The Naval Observatory Merged Astrometric Dataset (NOMAD) gives B and V magnitude of the object as 16.31 and 14.89 respectively with an error of ${\pm}$ 4.0 magnitude.
\end{table*}
}

\section{Radial velocity and stellar atmospheric parameters} \label{section atmo para}
Radial velocity of J045 is calculated from the shift in the observed wavelength from the lab wavelength for clean and unblended lines and is found to be $-$65.1$\pm$0.05 km s$^{-1}$ (Table \ref{radial velocity}).
\cite{Jonsson_2020} estimated a radial velocity of ${\sim}$ $-$39.9 km s$^{-1}$ for this object comparing its   stellar spectrum with a grid of synthetic spectra. The estimated radial velocity of the object  obtained  using the LAMOST 1D pipeline  by \cite{Luo_2015} is ${\sim}$   $-$70.1 km s$^{-1}$.  Our estimate of   radial velocity ($-$65.1 km s$^{-1}$) derived using several clean unblended lines  on the spectra of  this object is distinctly different from the estimate of \cite{Jonsson_2020} and  closer to the estimate  of \cite{Luo_2015}.  The difference in the estimates of radial velocity  indicates that the object could possibly be  in a binary system.  

{\footnotesize
\begin{table*}
\caption{\bf Radial velocities of the programme star. } \label{radial velocity}
\begin{tabular}{lcc}
\hline
Star            & V$_{r}$             & Reference \\
                & (km s$^{-1}$)          & \\
\hline
LAMOST J045019.27+394758.7  	&	 $-65.1$$\pm$0.05 	& 1	\\ 
                                &    $-70.1$            & 2 \\
                                &    $-39.9$$\pm$0.01   & 3 \\
\hline
\end{tabular}

1. Our work, 2. \cite{Luo_2015}, 3.\cite{Jonsson_2020}

\end{table*}

}

Atmospheric parameters, the effective temperature T$\rm_{eff}$, surface gravity, log\,g and microturbulent velocity ${\zeta}$ of J045 are estimated  using the equivalent widths measured  for unblended and clean lines of   Fe I and Fe II lines (Table A1). We have made use of  thirty four Fe I and two Fe II lines for our analysis. The line information are taken from linemake \footnote{linemake contains 
laboratory atomic data (transition probabilities, hyperfine and isotopic substructures) 
published by the Wisconsin Atomic Physics and the Old Dominion Molecular Physics groups. 
These lists and accompanying line list assembly software have been developed by 
C. Sneden and are curated by V. Placco at \url{https://github.com/vmplacco/linemake}.} 
\citep{Placco_2021}, an atomic and molecular line database.  

\begin{figure}
\centering
\includegraphics[width=9cm,height=9cm]{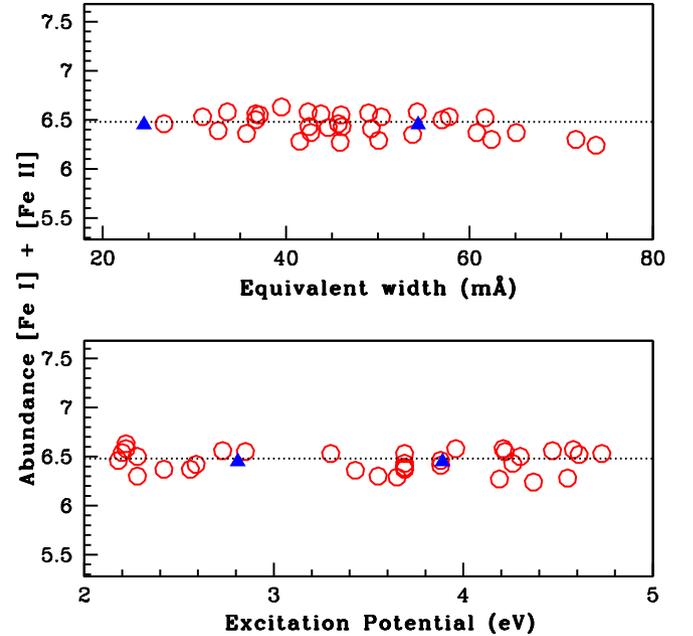} 
\caption{   The iron abundances of J045 are plotted  as a function of equivalent widths (upper panel) and excitation potential (lower panel). The open circles and the solid triangles represent the derived  abundance estimates for Fe I   and  Fe II lines respectively.} \label{ew ep plot}
\end{figure}

The effective temperature is  fixed at the value by an iterative process for which the trend between the abundances derived using  Fe I lines  and the  excitation potentials corresponding to those  lines gives a  slope close to zero. At this temperature, microturbulent velocity is adopted to be that value for which the 
Fe I lines abundances and the reduced equivalent widths do not exhibit any trends. At these values of effective temperature and microturbulent velocity, the surface gravity log g is fixed at the value, for which the abundances of Fe I and Fe II are nearly the same. The iron abundances of J045  as a function of excitation potential and as a function of equivalent widths are shown in Figure \ref{ew ep plot}.  The detailed procedure followed can be found  in our earlier papers \cite{Purandardas_2019a, Shejeelammal_2021a}. 
We used MOOG (\citealt{Sneden_1973}, updated version 2013) for our analysis considering  
local thermodynamic equilibrium. 
We have  selected the model atmospheres from the Kurucz grid of model atmospheres with 
no convective overshooting (\url{http://kurucz.harvard.edu/grids.html}). Solar abundances are taken
from \cite{Asplund_2009}.  The derived atmospheric parameters for J045 are presented in 
Table \ref{atm parameters}.

{\footnotesize
\begin{table*}
\caption{\bf LAMOST J045019.27+394758.7: Derived atmospheric parameters  and literature values. } \label{atm parameters}
\resizebox{\textwidth}{!}{\begin{tabular}{lccccccccccc}
\hline
T$_{eff}$  & log g  &$\zeta$ & [Fe I/H] & [Fe II/H] & [Fe/H] &  M$_{bol}$ & log(L/L$_{\odot}$) & Mass(M$_{\odot}$) & log g & Age  & Reference\\
(K)        & cgs    &(km s$^{-1}$)  &    &          &       &             &      &      &   (Parallax method) &        &\\
($\pm$100) & ($\pm$0.2)  & ($\pm$0.2)&   &          &       &             &      &    & (cgs) &Gyr   & \\
\hline	
4850	&	2.50	  &	0.75	  &	 $-$1.05$\pm$0.11 (34)	&	 $-$1.05$\pm$0.00 (2)	& $-$1.05 & $-0.92\pm$0.55 & 2.27$\pm$0.22 & 2.00$\pm$0.50 & 2.17$\pm$0.10 &  0.79$\pm$0.49 &  1\\ 
4627    &   2.26      & 0.72      &                        &                  & $-0.90$ &  & & & & &   2 \\ 
4250    &   1.00      & 1.55      &  &  & $-0.50$ & & & & & & 3 \\
\hline
\end{tabular}}

1. Our work, 2. \cite{Xiang_2019}), 3. \cite{Jonsson_2020}
\end{table*}
} 

\subsection{Mass and age} 

\begin{figure}
\centering
\includegraphics[width=\columnwidth]{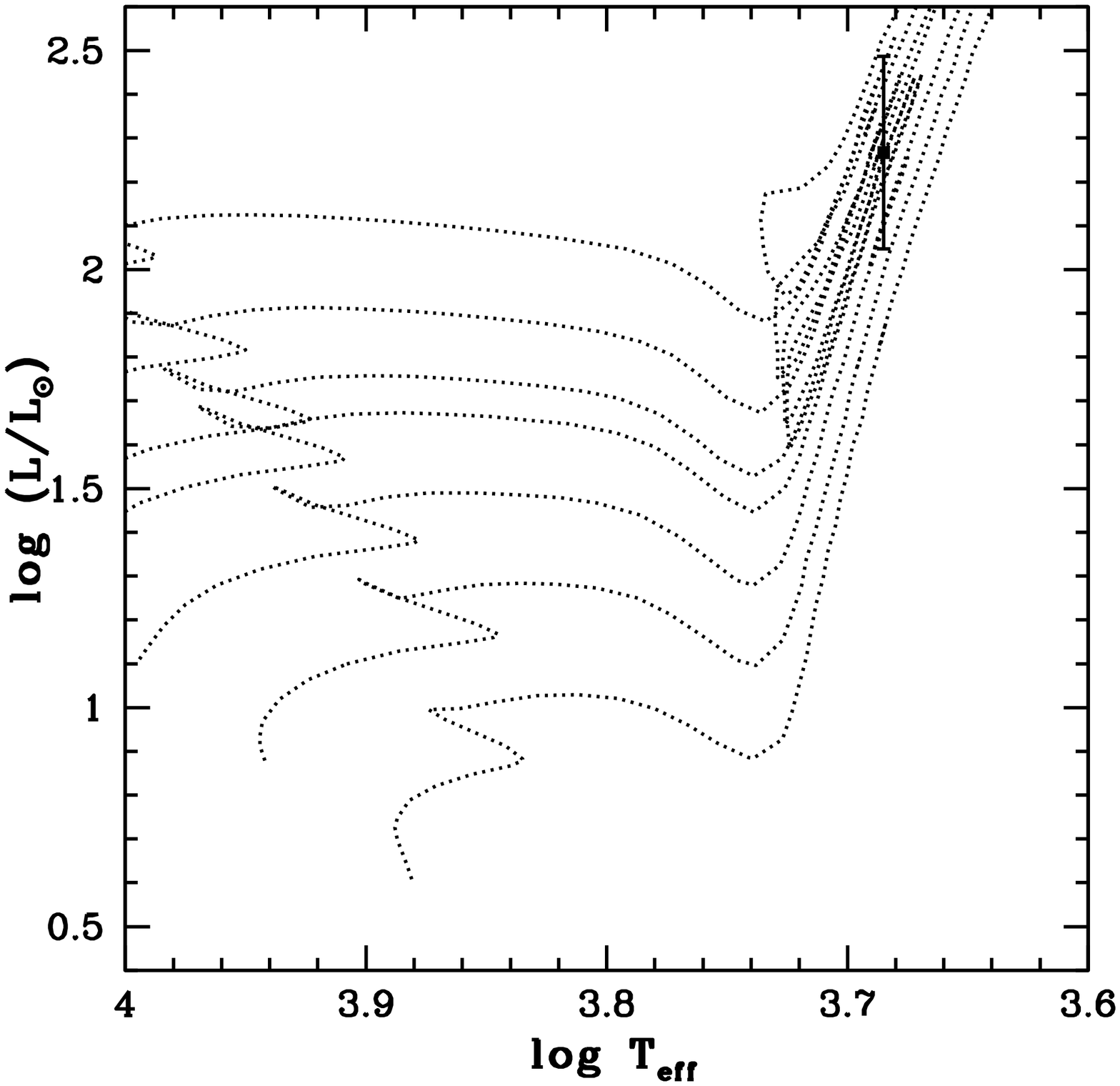}
\includegraphics[width=\columnwidth]{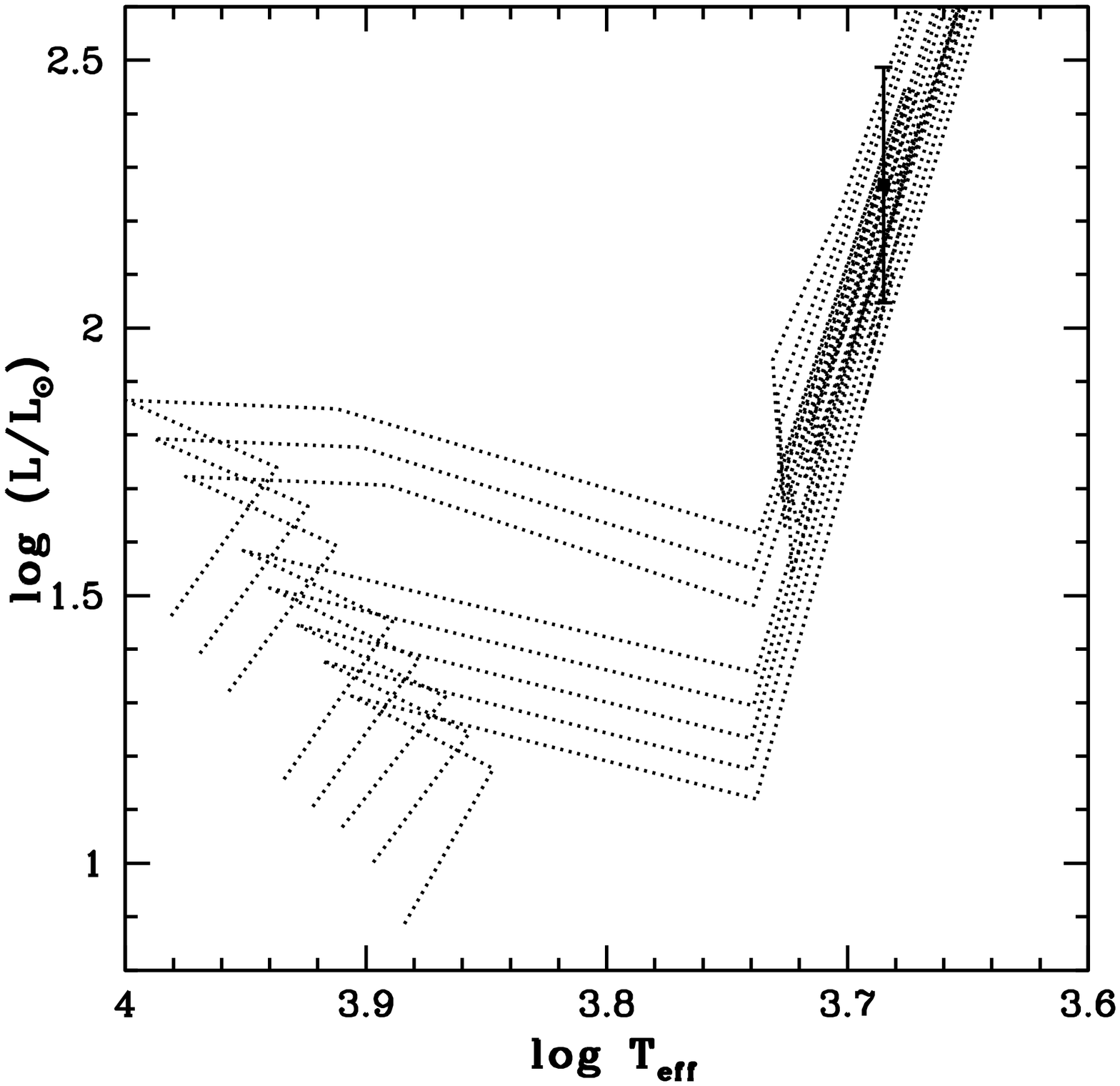}
\caption{The location of J045 on the H-R diagram is shown. The evolutionary tracks for 1.3, 1.5, 1.7, 1.9, 2.0, 2.2, and 2.5 M$_{\odot}$ are shown  in the upper panel from the  bottom to top. The isochrones for log(age) 9.25, 9.20, 9.15, 9.05, 8.95, 8.90, 8.85, and 8.80 are shown  in the bottom panel from the bottom to top. The age of the programme star corresponds to the isochrone for log(age) = 8.90. } \label{track}
\end{figure}

We could determine the mass and the age of J045 from its location on the H-R diagram. log(L/L$_{\odot}$) vs. T$\rm_{eff}$ (Figure \ref{track}). Luminosity of the star is determined using the relation, 
 \[  log(L/L_{\odot}) = (M_{\odot}-M_{bol})/2.5\] 
Here M$_{\odot}$ represents the  Sun's bolometric magnitude, and 
\[M_{bol} = Mv+BC-Av\].
Mv is determined using the equation, 
\[Mv = V-(5log(d))+5\]
The  visual magnitude V of the star is adopted from the Naval Observatory Merged Astrometric Dataset (NOMAD) and the parallax values are adopted from \textit{Gaia} (\citealt{Gaia_2016, Gaia_2018}, \url{https://gea.esac.esa.int/archive/}). Bolometric corrections are estimated using the empirical calibrations of \cite{Alonso_1999}.
Instellar extinction used for the determination of bolometric magnitude is estimated from the formula 
given in \cite{Chen_1998}. We have used the evolutionary tracks and the isochrones from \cite{Girardi_2000} corresponding to Z = 0.004 to determine 
the mass and the age of the star. From the location of J045 on the H-R diagram, we find that the object lies on the giant branch  with a mass $\sim$ 2 M$_{\odot}$. The estimated age of the object J045 is 0.79 Gyr. Estimates of the mass and the age from the parallax method are tabulated in Table \ref{atm parameters}.

\subsection{log g from parallax method}
The surface gravity log\,{g}  of the object J045 is derived  adopting our mass estimate in the relation  \\
log(g/g$_{\odot}$) = log(M/M$_{\odot}$) + 4log(T$_{eff}$/T$_{eff\odot}$) + 0.4(M$_{bol}$ $-$ M$_{bol\odot}$)\\
The adopted Solar parameters  log g$_{\odot}$ = 4.44, T$_{eff\odot}$ = 5770K, and M$_{bol\odot}$ = 4.75 mag are taken from \cite{Yang_2016}. The value of log g estimated using the parallax method is 2.17$\pm$0.10 cgs, which is slightly lower than the estimated  spectroscopic log\,{g} value. However, we have used the  estimated spectroscopic log\,{g} value throughout  our analysis.

\section{Determination of elemental abundances } \label{section abundance deter}
Elemental abundances  are estimated from the  equivalent widths measured for a number of  good lines due to
various elements. We
have also performed  spectrum synthesis calculations  whenever found applicable. Various lines  are identified from the overplot of the Arcturus spectrum on  the spectrum of J045.  A master line list was generated including
the equivalent width measurements, lower excitation potential of the lines and log\,{gf} values obtained using linemake.  

 We have used only clean and  symmetric  lines for the measurement of  equivalent widths.  We could  estimate the abundances for 25 elements as presented in the Table \ref{abundance}.
 We have performed  the spectrum synthesis calculation for the elements that show hyperfine splitting, 
 such as, Sc, V, Mn, Co, Ba, La, and Eu.  

\par In the Table \ref{abundance}, we have presented the abundance results. The lines used for the determination of elemental abundances are tabulated in  Table A2. Before presenting our discussion on elemental abundances, we present in the following, a discussion on verification of abundance estimates.

{\footnotesize
\begin{table*}
\caption{\bf Elemental abundances in LAMOST J045019.27+394758.7  } \label{abundance}
\begin{tabular}{lccccccccc}
\hline                      
 & Z & solar log$\epsilon^{\ast}$ & log$\epsilon$ & [X/H] & [X/Fe] \\
\hline
C (C$_{2}$, 5165 {\rm \AA}) & 6 & 8.43 & 7.49$\pm$0.20(syn)   & $-$0.94 & 0.11 \\
C (C$_{2}$, 5635 {\rm \AA}) & 6 & 8.43 & 7.49$\pm$0.20(syn)   & $-$0.94 & 0.11 \\
N         & 7    & 7.83    & 7.39$\pm$0.02(syn)  & $-$0.44 & 0.61 \\
O         & 8    & 8.69    & 7.51$\pm$0.20(syn)         & $-$1.18 & $-$0.13 \\
Na I      & 11   & 6.24    & 5.87$\pm$0.11(3)           & $-$0.37 & 0.68  \\
Mg I      & 12   & 7.60    & 6.63$\pm$0.20(2)           & $-$0.97 & 0.08  \\
Si        & 14   & 7.51    & 6.64$\pm$0.05(7)           & $-$0.87 & 0.18 \\
Ca I      & 20   & 6.34    & 5.31$\pm$0.09(7)           & $-$1.03 & 0.02 \\
Sc II     & 21   & 3.15    & 1.82(3,syn)                & $-$1.33 & $-$0.28 \\
Ti I      & 22   & 4.95    & 3.65$\pm$0.12(4)               & $-$1.30 & $-$0.25 \\
Ti II     & 22   & 4.95    & 3.65$\pm$0.14(5)               & $-$1.30 & $-$0.25 \\
V I       & 23   & 3.93    & 3.39$\pm$0.09(2,syn)       & $-$0.54 & 0.51 \\
Cr I      & 24   & 5.64    & 4.46$\pm$0.14(6)               & $-$1.18 & $-$0.13 \\
Mn I      & 25   & 5.43    & 4.33$\pm$0.10(2,syn)       & $-$1.10 & $-$0.05 \\
Fe I      & 26   & 7.50    & 6.45$\pm$0.11(34)              & $-$1.05 & -\\
Fe II     & 26   & 7.50    & 6.45(2)                       & $-$1.05 & -\\
Co I      & 27   & 4.99    & 4.06$\pm$0.06(3,syn)       & $-$0.93 & 0.12 \\
Ni I      & 28   & 6.22    & 5.31$\pm$0.13(12)                    & $-$0.91 & 0.14 \\
Zn I      & 30   & 4.56    & 2.89$\pm$0.11(2)                    & $-$1.67 & $-$0.62 \\
Sr I      & 38   & 2.87    & 1.92$\pm$0.20(1,syn)       & $-$0.95 & 0.10$^{a}$  \\
Y I       & 39   & 2.21    & 1.21(2,syn)                & $-$1.00 & 0.05 \\
Y II      & 39   & 2.21    & 1.28$\pm$0.14(2,syn)       & $-$0.93 & 0.12 \\
Ba II     & 56   & 2.18    & 1.04$\pm$0.04(2,syn)       & $-$1.14 & $-$0.09 \\
La II     & 57   & 1.10    & 0.18$\pm$0.04(2,syn)       & $-$0.92 & 0.13 \\
Ce II     & 58   & 1.58    & 0.39$\pm$0.14(3)               & $-$1.19 & $-$0.14 \\
Pr II     & 59   & 0.72    & $-$0.23$\pm$0.04(2,syn)              & $-$0.95 & 0.10 \\
Nd II     & 60   & 1.42    & 0.20$\pm$0.08(2, syn)               & $-$1.22 & $-$0.17 \\
Sm II     & 62   & 0.96    & $-$0.10(2,syn)                 & $-$1.06 & $-$0.01 \\ 
Eu II     & 63   & 0.52    & $-$0.24$\pm$0.11(2,syn)    & $-$0.76 & 0.29 \\
\hline
\end{tabular}

$^\ast$  \cite{Asplund_2009}.  The number of lines used for the abundance determination is given inside the parenthesis. \\
$^{a}$ After non-LTE correction

\end{table*}

}

\subsection{Verification of abundance estimates}
In order to check how  accurate our estimates are, we have performed an abundance analysis for the normal giant HD 111721 which has similar atmospheric parameters as that of J045. Two high-resolution spectra of this object, one from ELODIE  (\url{http://atlas.obs-hp.fr/elodie/}) at a resolution of R ${\sim}$ 42,000  and one from the SUBARU Archive (\url{http://jvo.nao.ac.jp/portal/v2/}) at a resolution R ${\sim}$ 50,000,  are used for our analysis.   The atmospheric parameters for HD 111721 (T$_{eff}$ = 4947 K,  log g = 2.63,  [Fe/H] = $-$1.34),   are adopted from \cite{Ishigaki_2012, Ishigaki_2013}. Our estimated elemental abundances when compared with those derived by \cite{Ishigaki_2012, Ishigaki_2013} are found to be in close agreement (differences less than 0.05 dex) as  shown in Table \ref{hd111721 abundance}. This agreement confirms reliability of our abundance estimates and the line information used for the elemental abundance calculations.   

{\footnotesize
\begin{table*}
\caption{\bf Comparison of the abundances of the star HD~111721 with the literature values. } \label{hd111721 abundance}
\resizebox{\textwidth}{!}{\begin{tabular}{lcccccccccc}
\hline
 Star name                    & [Na I/Fe]   & [Mg I/Fe]  & [Si I/Fe]  & [Ca I/Fe]   & [Sc II/Fe]  & [Ti I/Fe]  & [Ti II/Fe]  & [V I/Fe]  &  [Cr I/Fe]  & 
 [Cr II/Fe]  \\
\hline
 HD~111721                    & 0.00  & 0.40    & 0.34    & 0.31    & 0.20     & 0.28    & 0.40     & $-$0.09  & $-$0.14   & 0.17  \\
                            & 0.02  & -      & -      & -      & 0.22     & -      & -       & -       & $-$0.16   & 0.18  \\ 
                            & -    & 0.37    & 0.37    & 0.32    & -       & 0.26    & 0.42     & $-$0.06  & -        & - \\
\hline
                      & [Mn I/Fe]  & [Ni I/Fe]  & [Zn I/Fe]  & [Y II/Fe]  & [Ba II/Fe]  & [La II/Fe]  & [Nd II/Fe]  & [Sm II/Fe]  & [Eu II/Fe]   & Ref  \\
\hline
                    & $-$0.20    & $-$0.07   & 0.17  & 0.11   & 0.21     & 0.26   & 0.19   & 0.28  & 0.23   & 1  \\
                    &  $-$0.20   & $-$0.05   & 0.21  & 0.13   & 0.21     & 0.30   & 0.20   & 0.27  & 0.25   & 2  \\
                    & -         & -        & -    & -     & -       & -     & -     & -    & -     & 3  \\ 
\hline
\end{tabular}}

References : 1. Our work, 2. \cite{Ishigaki_2013}, 3. \cite{Ishigaki_2012}
 \end{table*}
}

\subsection{O, C and N}
\begin{figure}
\centering
\includegraphics[width=\columnwidth]{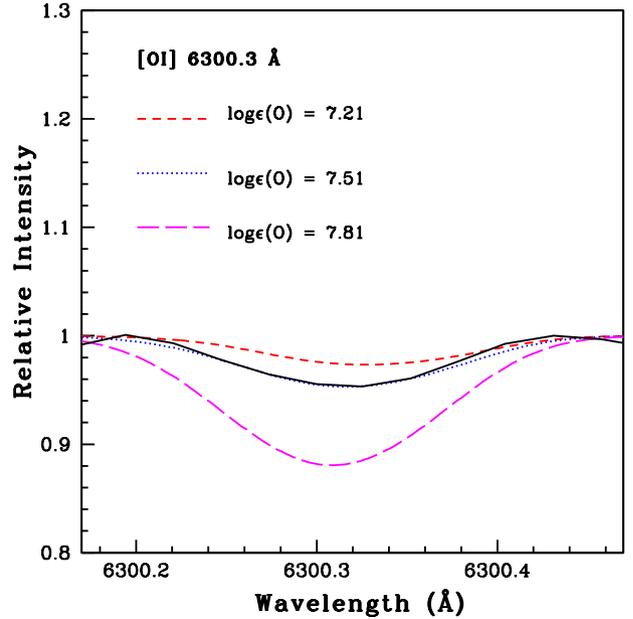}
\caption{ Synthesis of [OI] line around 6300 {\rm \AA}.  Observed and synthesized spectra are represented by solid and dotted  lines respectively.  The Short-dashed line is used to represent the synthetic spectrum corresponding to $\Delta$[O/Fe] = -0.3 and long-dashed line is used for  $\Delta$[O/Fe] = +0.3.} \label{OI synth}
\end{figure}

Spectrum synthesis calculation of the oxygen forbidden line [OI] 6300.3 {\rm \AA} (Figure \ref{OI synth}) is used to determine the oxygen abundance and  is found to be marginally under-abundant with [O/Fe] $\sim$ $-0.13$. We could not use the other lines of O such as [O I] 6363.7 {\rm \AA} and the oxygen triplets around 7774 {\rm \AA} as these lines are very weak and blended.

The [O I] forbidden line at 6300 {\rm \AA} is found to be influenced by Ni I blend.  However, the effect of Ni I blend decreases and becomes negligible in stars with metallicity $<$ $-$1 \citep{Prieto_2001}.  

\begin{figure}
\centering
\includegraphics[width=\columnwidth]{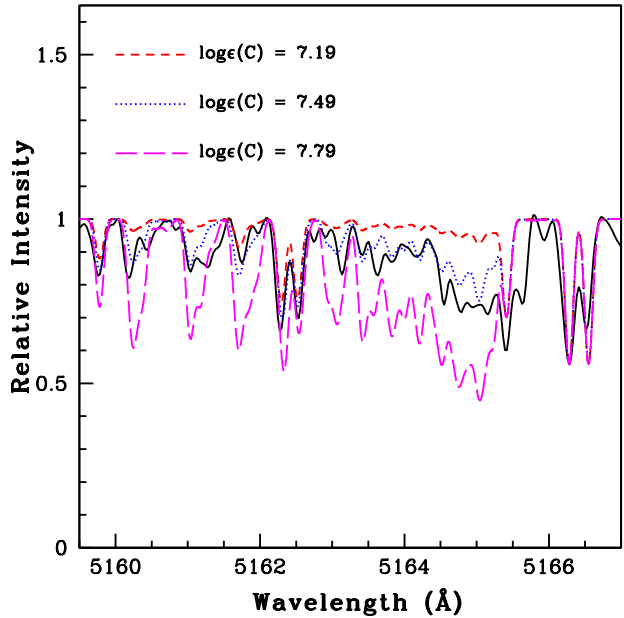}
\caption{Synthesis of C$_{2}$ band around 5165 {\rm \AA}. 
 Observed and synthesized spectra are represented by solid and dotted  lines respectively.
The short-dashed line is used to  represent the synthetic spectrum corresponding to $\Delta$ [X/Fe] = -0.3 and the long-dashed line for  $\Delta$[X/Fe] = +0.3.} \label{carb synth}
\end{figure}

The spectrum synthesis  of C$_{2}$ bands at 5165 and 5635 \r{A} (Figure \ref{carb synth}) is used to estimate the abundance of carbon which gives [C/Fe] $\sim$ 0.11. We could not use CH band for the abundance determination of carbon as this region is  very noisy. The spectrum synthesis of $^{12}$CN lines at 
8003, 8003.5 and 8004 {\rm \AA} are used for the estimation of nitrogen abundance.
 The three $^{12}$CN lines give similar values for [N/Fe], 0.60, 0.63, and 0.59 respectively. The average [N/Fe] = 0.61 is listed in Table \ref{abundance}. 
 
\par
\begin{figure}
\centering
\includegraphics[width=9cm,height=9cm]{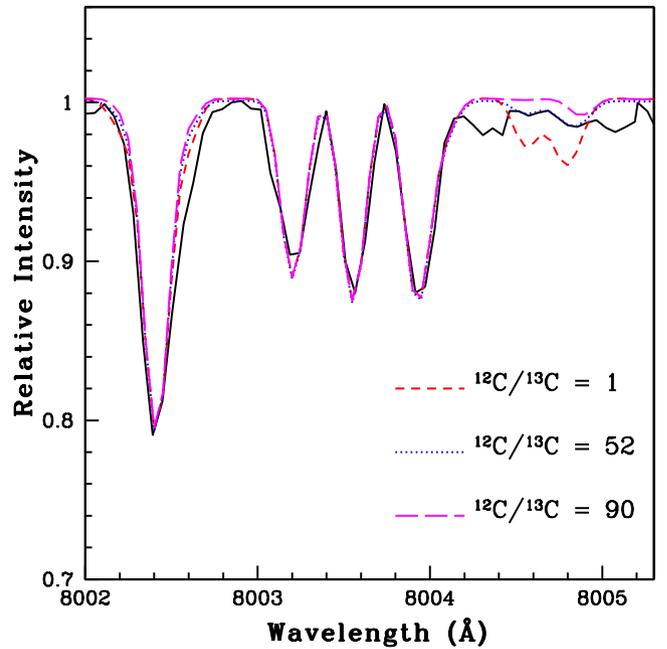}
\caption{The spectral synthesis fits of CN band around 8005 {\rm \AA} are shown in this figure. The observed spectrum is represented by solid line.  The dotted line represents the best fit synthetic spectrum corresponding to $^{12}$C/$^{13}$C = 52. The short- and the long-dashed lines  illustrate the sensitivity of the line strengths to the isotopic carbon abundance ratios.} \label{cir synth}
\end{figure}

 Using the estimated nitrogen and carbon abundances, we have determined the carbon isotopic ratio  from the spectrum synthesis calculation of the CN band at 8005 \r{A} (Figure \ref{cir synth}). The 
estimated  $^{12}$C/$^{13}$C $\sim$ 52 and  [C/N] = $-$0.50.

\subsection{Odd-Z element Na}
The equivalent widths of Na I lines at 5682.633, 6154.226, and 6160.7 \r{A} are used for the estimation of Na abundance. Na is moderately enhanced in J045  with [Na/Fe] $\sim$ 0.68. 

\subsection{ Mg, Si, Ca, Sc, Ti, V}
The equivalent widths  of Mg I 4730.029, 5711.088 \r{A} lines are used for the determination of Mg abundance which give an average value of [Mg/Fe] $\sim$ 0.08. Silicon abundance is determined from the equivalent widths  of seven Si I lines (Table A2) which gives a value of [Si/Fe] $\sim$ 0.18. For the estimation of Ca abundance we have used a few Calcium I lines as presented in the Table A2 that give near-solar Ca abundance.  The abundance of scandium is estimated from the spectrum synthesis calculations of Sc II 6245.637, 6309.920, 6604.601 \r{A} lines which give an average value of [Sc/Fe] $\sim$ $-$0.28. 
The  equivalent widths of four Ti I and five Ti II lines (Table A2) are used for the estimation of 
Ti abundance. Our estimation shows  Ti is  slightly under abundant with [Ti I, Ti II/Fe] $\sim$ $-$0.25.
Spectrum synthesis of V I lines at 5727.048\r{A} and 6251.827\r{A} give  log${\epsilon}$(V) values as 3.32 and 3.45 respectively. In Table \ref{abundance}, an average  value of [V/Fe] $\sim$ 0.51 is listed.

\subsection{Cr, Mn, Co, Ni, Zn}
The equivalent widths of a few Cr I lines  (Table A2) are used to derive the Cr abundance. Cr is found to be marginally under-abundant with [Cr/Fe] $\sim$ $-$0.13. Spectrum synthesis of the Mn I lines 
6013.51, 6021.89 \r{A} are used to estimate the abundance of Manganese   which give an average value of [Mn/Fe] $\sim$ $-$0.05.
The spectrum synthesis calculations of Co I 5342.695, 5483.344, 6632.4 \r{A} lines are used to determine the abundance of Co which is found to be marginally enhanced. 
Abundances of nickel and zinc are estimated using  the equivalent widths measured for  twelve Ni I  lines
and two Zn I lines 4722.2 \AA\, and 4810.5 \AA\, (Table A2). These two lines returned log${\epsilon}$(Zn) values as 2.97 and 2.81 respectively.  The average  LTE abundance estimates give  [Zn/Fe] = $-$0.62.
 The S/N (=57 ) ratio at 4750 \AA\ is  reasonably sufficient for reliable abundance measurements. While the abundance of Ni is found to be marginally enhanced,  we find that the Zn is under abundant with [Zn/Fe] $\sim$ $-$0.62. 
    We have not applied non-LTE
 correction to the derived Zn abundance. The
 corrections for non-LTE effects on the two Zn  lines used for abundance analysis
  are shown  to be positive and relatively  insignificant 
  ($<$  0.1 dex)  in the range of metallicity  $-$4.0 ${\le}$ [Fe/H] ${\le}$ 0 \citep{Takeda_2005}.

\subsection{Sr, Y, Zr}
The abundances of  strontium and yttrium in J045 are found to be close to the solar values. 
The spectral synthesis calculation of Sr I 4607.327 \r{A} is used to derive the Sr abundance. We have applied the non-LTE correction to the estimated Sr abundance as given by \cite{Bergemann_2012}. For a star with T$\rm_{eff}$ = 4800, log g = 2.20, and [Fe/H] $\sim$ $-$1.20, non-LTE correction is 0.32 \citep{Bergemann_2012}.  Figure \ref{ba sr synth} (left panel) shows the spectrum synthesis fits for the
line Sr I 4607.3 \r{A}. The abundance of yttrium is derived  from  the spectrum synthesis calculation of Y I 5630. 130, 6435.004 \r{A} lines and Y II 5119.112, 5402.774 \r{A} lines.  Due to the absence of good 
usable lines of zirconium in the spectrum of J045 we could not estimate Zr abundance. 
\subsection{Ba, La, Ce, Pr, Nd, Sm, Eu}
\begin{figure}
\centering
\includegraphics[width=\columnwidth]{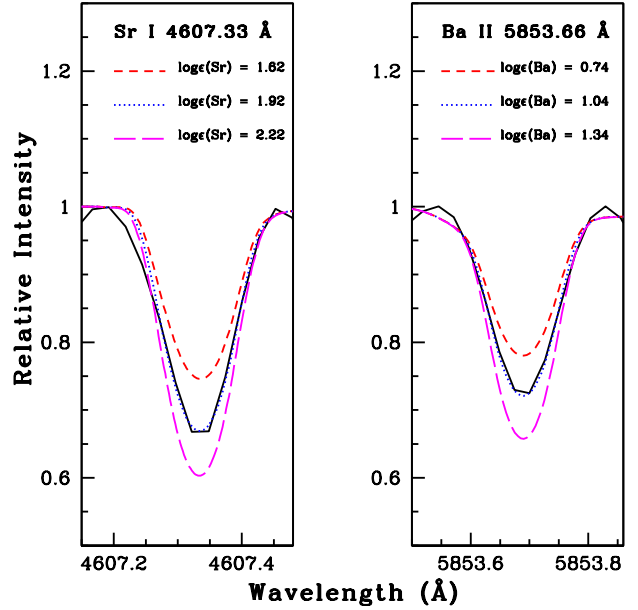} 
\caption{Spectral synthesis fits of the lines Sr I 4607.33 {\rm \AA} (left panel) and Ba II 5853.66 {\rm \AA} (right panel). 
The dotted lines represent synthesized spectra and solid lines indicate the observed spectra. The short-dashed lines represent the synthetic spectra corresponding to $\Delta$ [X/Fe] = -0.3 and the
long-dashed lines correspond to $\Delta$[X/Fe] = +0.3. 
} \label{ba sr synth}
\end{figure}

The estimated abundances of Ba, Pr, and Sm are found to be close to solar values.  La is marginally enhanced, Ce and Nd are found to be
marginally under-abundant, and Eu is only slightly enhanced.
The spectrum syntheses of  Ba II 5853.668, 6141.713 \r{A} lines (Figure \ref{ba sr synth}, right panel)
return near-solar values  with  [Ba/Fe] $\sim$ $-$0.09. Spectrum synthesis calculations of the lines La II 4921.776, 5303.528 \r{A} are used to determine the abundance of lanthanum which give the value [La/Fe] $\sim$ 0.13. 
Abundance of cerium is calculated using  equivalent widths of three lines (Table A2) and is found to show the value [Ce/Fe] $\sim$ $-$0.14. The spectral synthesis calculations of Pr II 5259.728, 6165.891 \r{A} lines are used to find the 
Pr abundance and the spectral synthesis calculations of Nd II 4797.153, 5319.810 \r{A} lines are used to determine the  neodymium abundance. The estimated abundances of Pr and Nd  are [X/Fe]$\sim$0.10, $-$0.17 respectively. 
 The spectrum synthesis calculation of the lines Sm II 4566.200, 4676.9 \r{A}  give a value [Sm/Fe] ${\sim}$ $-$0.01. The spectrum synthesis calculations of the lines Eu II 6437.640, 6645.064 \r{A} 
 return a slightly enhanced  europium abundance  with [Eu/Fe]$\sim$0.29.

\par We have calculated  the values of [ls/Fe], [hs/Fe], [hs/ls] ratios, where ls indicates Sr and Y the two light s-process elements,  and the hs indicates  Ba, La, Ce, and Nd, the heavy s-process elements. These values are presented in Table \ref{hs ls ratio}. A comparison between the  estimated abundances of 
J045  and the literature values  (whenever available)  is presented in Table \ref{literature comparison}.

{\footnotesize
\begin{table*}
\caption{\bf Estimates of [Fe/H], [ls/Fe], [hs/Fe], [hs/ls], $^{12}$C/$^{13}$C and [C/N]} \label{hs ls ratio}
\begin{tabular}{lccccccc}
\hline                       
Star name    & [Fe/H]  & [ls/Fe] & [hs/Fe] & [hs/ls] & $^{12}$C/$^{13}$C & [C/N] \\ 
\hline
LAMOST J045019.27+394758.7 & $-$1.05 & 0.08 & $-$0.07 & $-$0.15 &  52 $\pm $ 3.0 & $-$0.50 $\pm$ 0.32 \\
\hline
\end{tabular}

\end{table*}
}

{\footnotesize
\begin{table*}
\caption{\bf A comparison of the abundances of the programme star with  literature values. } \label{literature comparison}
\resizebox{\textwidth}{!}{\begin{tabular}{lccccccccc}
\hline
Star name                   & [Fe/H]  & [C/Fe] & [N/Fe] & [O/Fe]  & [Na/Fe] & [Mg/Fe] & [Si/Fe] & [Ca/Fe] &  [Ti/Fe]\\
\hline
LAMOST J045019.27+394758.7  & $-$1.05 & 0.11   & 0.61   & $-$0.13 & 0.68    & 0.08    & 0.18    & 0.02 & $-$0.25  \\
                            & $-0.90$ & 0.47   & 0.47   & $-0.01$ & 0.98    & 0.04    & 0.04    & 0.51 & $-0.06$\\ 
                            & $-0.50$ & 0.54   & 0.18   & 0.04    & 0.15    & 0.10    & 0.04    & 0.09&$-0.13$\\
\hline
                    & [V/Fe]  & [Cr/Fe] & [Mn/Fe] & [Co/Fe] & [Ni/Fe] & [Ba/Fe] & [Ce/Fe] & Ref& \\
\hline
                    & 0.51    & $-$0.13 & $-$0.05 & 0.12    & 0.14    & $-$0.09 & $-$0.14    & 1 &\\
                    &  -      & 0.19    & $-0.21$ & $-0.02$ & 0.04    & $-0.44$ & -         & 2 &\\
                    & $-0.07$ & 0.12    & $-0.17$ & $-0.15$ & 0.00    & -       & 1.08      & 3 &\\ 
\hline
\end{tabular}}

References : 1. Our work, 2. \cite{Xiang_2019}, 3. \cite{Jonsson_2020}
 \end{table*}
}

\section{Abundance uncertainties} \label{section abundance uncertainty}
Random and systematic errors produce uncertainties in the estimated elemental abundances. Random errors are caused by the uncertainties in the line parameters like equivalent widths, blending of lines and oscillator strength. The systematic error is produced due to the uncertainties in the stellar atmospheric parameters adopted for the analysis. 

\par We have calculated these uncertainties of our  abundance estimates following the steps  as discussed in \cite{Shejeelammal_2021a}.
The total uncertainties in the abundance estimates,  log$\epsilon$ is given by \\

$\sigma^{2}_{log\epsilon} = \sigma^{2}_{ran}+ (\frac{\partial log\epsilon}{\partial T})^{2} \sigma_{T_{eff}}^{2}+ (\frac{\partial log\epsilon}{\partial logg})^{2}\sigma^{2}_{log g}+  (\frac{\partial log\epsilon}{\partial \zeta})^{2}\sigma^{2}_{\zeta}+(\frac{\partial log\epsilon}{\partial [Fe/H]})^{2}\sigma^{2}_{[Fe/H])}$\\

where $\sigma_{ran}$ = $\sigma_{s}/\sqrt{N}$ . $\sigma_{s}$ indicates the standard deviation in the abundance estimates which is calculated from the number of lines, N,  corresponding to that particular element. The $\sigma$’s represent the uncertainties in the adopted atmospheric parameters of the star, and are given by T$\rm_{eff}$ $\sim$ $\pm$ 100 K, log g $\sim$ $\pm$ 0.2 dex, $\zeta$ $\sim$ $\pm$ 0.2 km s$^{-1}$ , and [Fe/H] $\sim$ $\pm$ 0.1 dex. The uncertainty in [X/Fe] is calculated  using the relation :
\[\sigma^{2}_{[X/Fe]} = \sigma^{2}_{X} + \sigma^{2}_{[Fe/H]}\]
The differential elemental abundances obtained for  J045 is presented in Table \ref{differential abundance}.

{\footnotesize
\begin{table*}
\centering
\caption{\bf Derived differential elemental abundances (log$\epsilon$)  for the object LAMOST J045019.27+394758.7. } \label{differential abundance}
\begin{tabular}{lccccccc}
\hline                       
element	& $\Delta$ T$_{eff}$ &	$\Delta$ log g & $\Delta \zeta$ &	$\delta$ [Fe/H] &	($\sum$ $\sigma^{2}_{i}$)$^{1/2}$ &	$\sigma$[X/Fe]\\
 &($\pm$100K)&	($\pm$0.2dex)&	($\pm$0.2kms$^{-1}$) &	($\pm$0.1 dex) &  &\\	
\hline
C	&	$\pm$0.02	&	0.00	&	$\pm$0.01	&	$\mp$0.01	&	0.00	&	0.02	\\
N	&	$\pm$0.29	&	$\pm$0.05	&	$\pm$0.04	&	$\pm$0.13	&	0.11	&	0.32	\\
O	&	$\pm$0.02	&	$\pm$0.08	&	$\pm$0.02	&	$\pm$0.03	&	0.01	&	0.09	\\
Na I 	&	$\pm$0.07	&	$\mp$0.01	&	$\mp$0.02	&	$\mp$0.01	&	0.01	&	0.07	\\
Mg I	&	$\pm$0.06	&$\pm$0.01	&	$\pm$0.01	&	$\pm$0.02	&	0.00	&	0.06	\\
Si	&	0.00	&	$\pm$0.02	&	$\mp$0.01	&	$\pm$0.01	&	0.00	&	0.02	\\
Ca I	&	$\pm$0.09	&	$\mp$0.02	&	$\mp$0.06	&	$\mp$0.01	&	0.01	&	0.11	\\
Sc II	&	$\pm$0.01	&	$\pm$0.09	&	$\mp$0.01	&	$\pm$0.05	&	0.01	&	0.10	\\
Ti I	&	$\pm$0.17	&	0.00	&	$\mp$0.07	&	$\mp$0.01	&	0.03	&	0.18	\\
Ti II	&	0.00	&	$\pm$0.08	&	$\mp$0.02	&	$\pm$0.03	&	0.01	&	0.09	\\
V I	&	$\pm$0.17	&	0.00	&	$\mp$0.01	&	$\mp$0.02	&	0.03	&	0.17	\\
Cr I	&	$\pm$0.14	&	$\mp$0.01	&	$\mp$0.09	&	$\mp$0.01	&	0.03	&	0.17	\\
Fe I	&	$\pm$0.07	&	$\pm$0.02	&	$\mp$0.05	&	$\pm$0.01	&	0.01	&		\\
Fe II	&	$\mp$0.07	&	$\pm$0.07	&	$\mp$0.03	&	$\pm$0.05	&	0.01	&		\\
Mn I	&	$\pm$0.10	&	$\mp$0.02	&	$\mp$0.01	&	$\mp$0.01	&	0.01	&	0.10	\\
Co I	&	$\pm$0.10	&	$\mp$0.01	&	0.00	&	$\mp$0.02	&	0.01	&	0.10	\\
Ni I	&	$\pm$0.09	&	$\pm$0.02	&	$\mp$0.04	&	$\pm$0.01	&	0.01	&	0.10	\\
Zn I	&	$\mp$0.01	&	$\pm$0.05	&	$\mp$0.03	&	$\pm$0.02	&	0.00	&	0.06	\\
Sr I	&	$\pm$0.24	&	$\pm$0.01	&	$\mp$0.03	&	0.00	&	0.06	&	0.24	\\
Y II	&	$\mp$0.05	&	$\pm$0.02	&	0.00	&	$\pm$0.01	&	0.00	&	0.05	\\
Ba II	&	$\mp$0.03	&	$\mp$0.02	&	$\mp$0.04	&	$\mp$0.01	&	0.00	&	0.05	\\
La II	&	$\mp$0.04	&	$\pm$0.10	&	$\pm$0.02	&	$\pm$0.08	&	0.02	&	0.14	\\
Ce II	& $\pm$0.02	&	$\pm$0.08	&	$\mp$0.02	&	$\pm$0.03	&	0.01	&	0.09	\\
Pr II	&	$\pm$0.03	&	$\pm$0.08	&	$\mp$0.01	&	$\pm$0.03	&	0.01	&	0.09	\\
Nd II	&	$\pm$0.03	&	$\pm$0.09	&	$\mp$0.01	&	$\pm$0.04	&	0.01	&	0.10	\\
Sm II	&	$\pm$0.01	&	$\pm$0.10	&	$\mp$0.04	&	0.00	&	0.01	&	0.11	\\
Eu II	&	$\pm$0.00	&	$\pm$0.08	&	$\mp$0.01	&	$\pm$0.04	&	0.01	&	0.09	\\

\hline   
\end{tabular}
\end{table*}
}

\section{TESS photometry: search for variability} \label{section tess}
\begin{figure}
\centering
\includegraphics[width=\columnwidth]{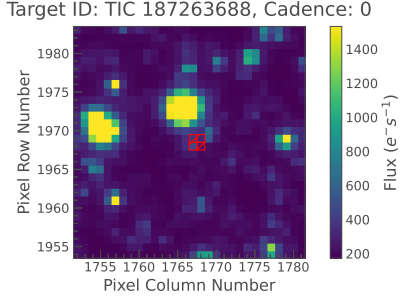}
\caption{ TESS Full Frame Image  (FFI). The object location and the aperture used is shown in red.} \label{TESS figure}
\end{figure}

\begin{figure}
\centering
\includegraphics[width=\columnwidth]{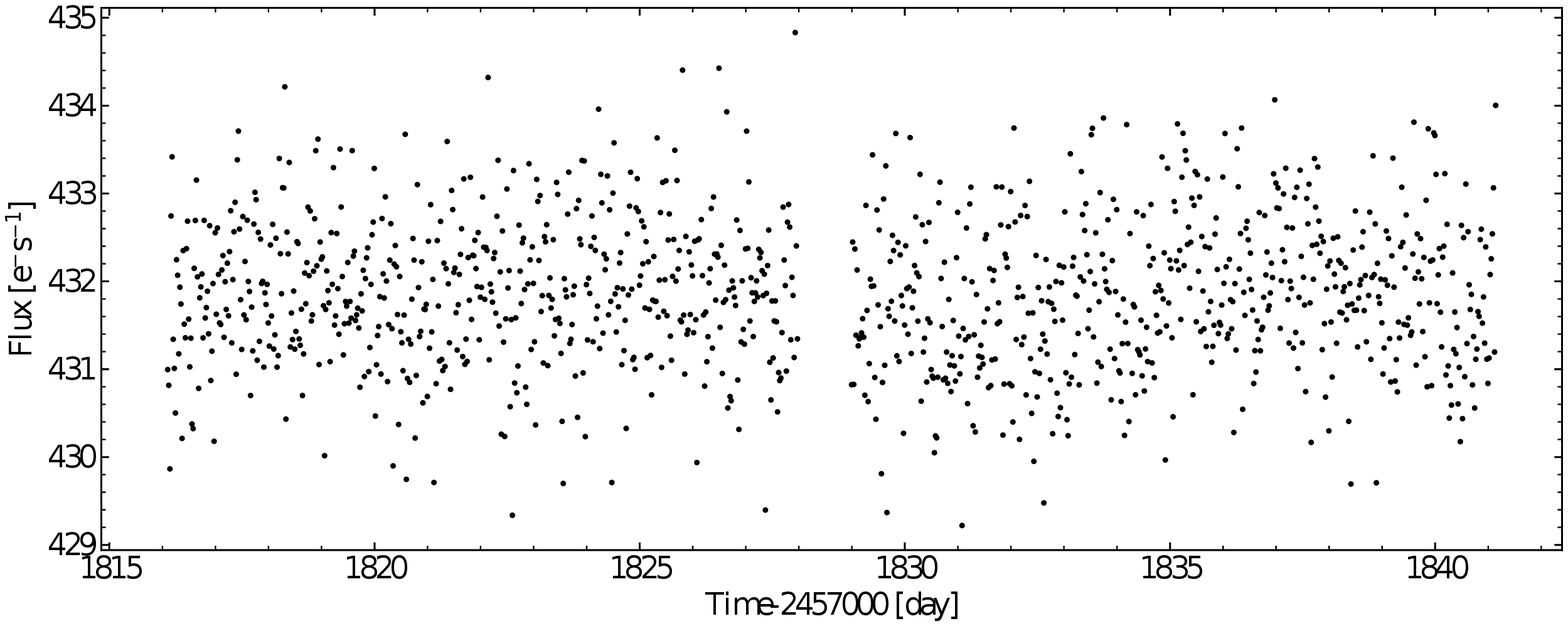}
\includegraphics[width=\columnwidth]{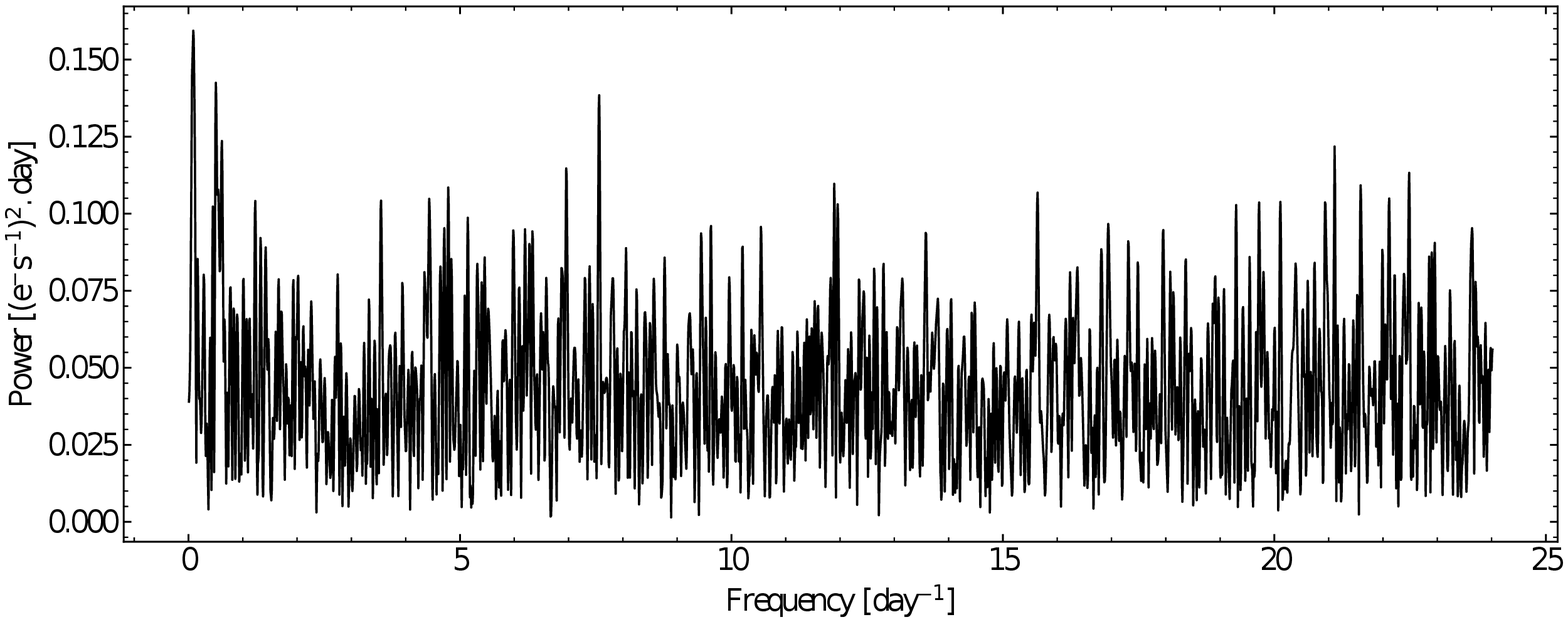}
\caption{ The regression corrected TESS light curves extracted from FFI 
(upper panel). Lomb-Scargle periodogram to check Fourier analysis of sinusoidal 
variations in the light curve (Bottom panel).} \label{TESS figure2}

\end{figure}

We have used TESS photometric data to search for variability in J045. The Transiting Exoplanet Survey Satellite (TESS), the first high-precision full-sky photometry survey in space observed this LAMOST target (TIC 187263688) from September 28, 2019  to December 23, 2019,  in sector 19 with Camera 1 CCD 3, exposure time of 1426s and  30-miniutes cadence. We acquried TESS Full Frame Images using Lightkurve \citep{2018ascl.soft12013L} which utilizes the TESSCut tool \citep{Brasseur_2019}. The target is near the limiting magnitude of TESS and  the data is quite noisy. Since our target is quite faint, we have created appropriate aperture as shown in Figure \ref{TESS figure}. To account for the instrumental and noise effects, we have applied regression correction to our light curve to remove trends, and set the light curve to the mean level. TESS scattered light is removed by subtracting model of the background that is built in Regression Correction. As seen from the Figure \ref{TESS figure2}, we do not see any signatures of variability. Nonetheless, we employed Lomb-Scargle algorithm commonly used for detecting and characterising periodic signals in un-evenly sampled data \citep{Lomb_1976, Scargle_1982}. Lomb-Scargle periodogram is shown in Figure \ref{TESS figure2}. The Lomb-Scargle algorithm performs computation of a Fourier-like power spectrum from un-evenly sampled data allowing determination of the period of oscillation. No period of oscillation could be detected from the limited TESS data set.

\section{Kinematic analysis} \label{section kinematic}
Kinematic analysis of J045 is performed following the detailed procedure as described in \cite{Purandardas_2021a}. The space velocity  is derived  using a right-handed coordinate system. In this system,  U, V and W the three components of the space velocity are  such that in the direction of the Galactic centre  U is positive,  V is positive in the direction of the  Galactic rotation and W is positive in the direction of the North Galactic Pole \citep{Johnson_1987}.  The values of parallax and proper motion  of the star are taken from \cite{Gaia_2016, Gaia_2018} and we have used our estimate of radial velocity for the calculation.

\par We have also checked whether the programme star belongs to the thin disk, thick disk or the halo population. We have followed the procedures as discussed in \cite{Reddy_2006, Bensby_2003, Bensby_2004, Mishenina_2004}, and  the assumption that the Galactic space velocity of the star has Gaussian distribution:

\[f(U,V,W) = K\times\exp[-\frac{U_{LSR}^{2}}{2\sigma _{U}^{2}}-\frac{(V_{LSR}-V_{asy})^{2}}{2\sigma _{V}^{2}}-\frac{W_{LSR}^{2}}{2\sigma _{W}^{2}}] \]

Where $K = \frac{1}{(2\pi) ^\frac{3}{2}\sigma _{U}\sigma _{V}\sigma _{W}}$\\
The values for characteristic velocity dispersion $\sigma _{U}$, $\sigma _{V}$  and $\sigma _{W}$ and the asymmetric drift $V_{asy}$ are adopted from \cite{Reddy_2006}. We have presented the results of the kinematic analysis of the programme star in Table \ref{kinematic analysis}. 
 \cite{Chen_2004},  have shown that  the thin disk objects have spatial velocity $<$ 85 km s$^{-1}$  and thick disk objects have spatial velocity in the range 85 - 180 km  s$^{-1}$.  Our estimated space velocity ${\sim}$ 78  km s$^{-1}$ therefore indicates that the object J045 to be a thin disk object. 
On the other hand, from high-resolution spectroscopic study of northern sky  dwarf stars,   \cite{Mikolaitis_2019}
have found that the stars with a lower combined velocity V$_{spa}$ $<$ 50 km s$^{-1}$
(V$_{spa}$     = (U$_{LSR}$+V$_{LSR}$+W$_{LSR}$)$^{1/2}$) are most probably the thin-disk stars, whereas those with 
v$_{spa}$ $>$ 50 km s$^{-1}$ should belong to the thick-disk.  Although following  \cite{Mikolaitis_2019}’s  definition J045 is likely a thick disk object, if we consider the criteria that all stars with p$_{thick}$/p$_{thin}$ $>$ 2 are thick-disk candidates and stars with p$_{thick}$/p$_{thin}$ $<$ 0.5 are thin disk objects \citep{Bensby_2003, Bensby_2004}, then, the J045 with p$_{thick}$/p$_{thin}$ $<$ 0.5 seems to belong to thin-disk population.

{\footnotesize
\begin{table*}
\caption{\bf Spatial velocity and probability estimates for the programme star} \label{kinematic analysis}
\resizebox{0.9\textwidth}{!}{\begin{tabular}{lcccccccc}
\hline                       
Star name           & U$_{LSR}$         & V$_{LSR}$           & W$_{LSR}$ & V$_{spa}$  & p$_{thin}$ & p$_{thick}$ & p$_{halo}$ & Population \\
                    & (kms$^{-1}$)        & (kms$^{-1}$)          & (kms$^{-1}$) &  (kms$^{-1}$) &           &           & &  \\  
    \hline
LAMOST J045019.27+394758.7 & 68.69$\pm$1.56      & $-29.92$$\pm$6.90    & $-22.07$$\pm$8.86      & 78.11$\pm$3.74    & 0.94 & 0.05 & 0.00 & Thin \\  
\hline
\end{tabular}}

\end{table*}
}

\section{Abundance Analysis: discussion} \label{section discussion}
The object J045 is included as a possible carbon star in LAMOST DR2 catalog,  however, 
our detailed analysis shows that this object does not show any spectral characteristics of carbon stars and  has a carbon abundance of near-solar value. The object is found to be a  metal-poor giant with metallicity [Fe/H] $\sim$ $-1.05$ with unusual abundance for several key elements.
\par While the abundances of C is near-solar, N is found to be high with [N/Fe] = 0.61.  One of the possible reasons for the observed enhancement of nitrogen in giant stars is the mixing which brings the CNO processed material to the surface that makes the atmosphere of the star "N rich" and "C poor" \citep{Spite_2005}. However, as discussed below, the program star does not show any signatures of mixing based on [C/N] and $^{12}$C/$^{13}$C ratios. However, if we consider the lower limit of error in [C/N] ratio, the star is well mixed. The origin of enhancement of nitrogen in a giant star can also be interpreted based on several scenarios. One such scenario is the accretion of N rich material from an AGB star \citep{Lennon_2003}. However, the observed abundance patterns in the program star does not support any AGB companion as the possible source of the nitrogen enhancement. Early accretion of Globular Clusters or dwarf spheroidal galaxies can also lead to nitrogen enhancement \citep{Fernandez_2019}. This scenario makes the star chemically distinguishable from the stars that belongs to disk, halo, or bulge population. The dynamical history of such stars is not yet clearly understood \citep{Fernandez_2019}. 

\par
\begin{figure}
\centering
\includegraphics[width=\columnwidth]{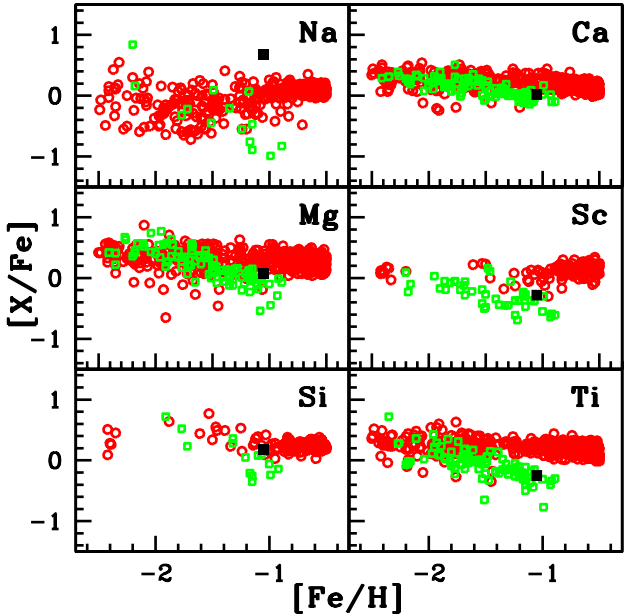}
\includegraphics[width=\columnwidth]{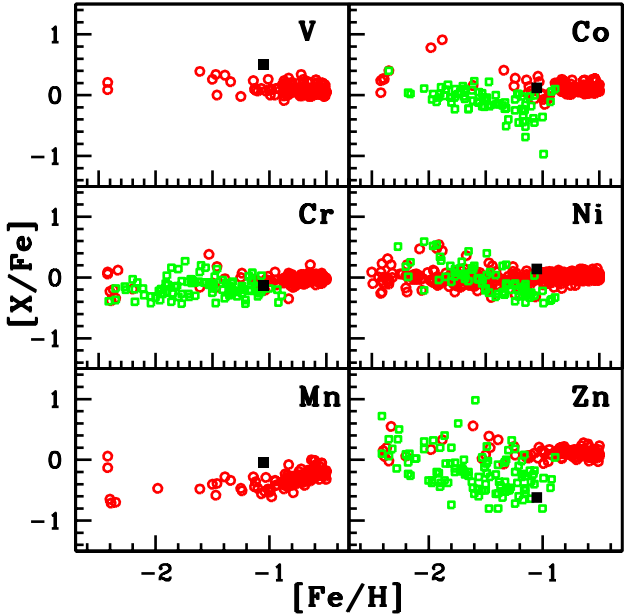}
\caption{ A comparison of the light elements abundance ratios with their counter parts observed in Galactic stars and Sculptor dwarf galaxy stars.  Comparisons are shown for elements for which literature data are available. Abundance ratios of  elements with respect to metallicity estimated in J045 are represented using filled square in black colour. Red open circles represent Galactic normal giants from literature
\citep{Honda_2004, Venn_2004, Aoki_2007, Luck_2007, Hansen_2016c, Purandardas_2019a}. Sculptor dwarf galaxy stars from literature \citep{Skuladottir_2017, Hill_2019} are represented using green open squares.} \label{light comparison}
\end{figure}

  O is found to be marginally under-abundant with [O/Fe] $\sim$ $-$0.13, 
a value which is highly unusual for a
Galactic star of metallicity [Fe/H] ${\sim}$ $-$1.0. None of the ${\alpha}$-elements Mg, Si and Ca are 
compatible with [${\alpha}$/Fe] - enhancement (i.e., 0.3 to 0.4 dex) characteristic of Galactic metal-poor stars of similar metallicity. J045 shows almost solar values with [Mg/Fe] = 0.08, [Si/Fe] = 0.18, and [Ca/Fe] = 0.02.   Such  abundance pattern is however compatible with stars of similar metallicity of Sculptor Dwarf Galaxy as shown in Figure \ref{light comparison} (upper panel).  This comparison provides observational evidence to argue that the object could be a possible Dwarf Galaxy 
escapee in the Galaxy.  Finding globular cluster (GC) escapees are not uncommon in the Galactic halo.  While \cite{Martell_2011} give  an estimate of GC escapees in the halo to be $\sim$ 3\% ,  an 
estimate of  Dwarf galaxy escapees is not available in the literature.

\par
\begin{figure}
\centering
\includegraphics[width=\columnwidth]{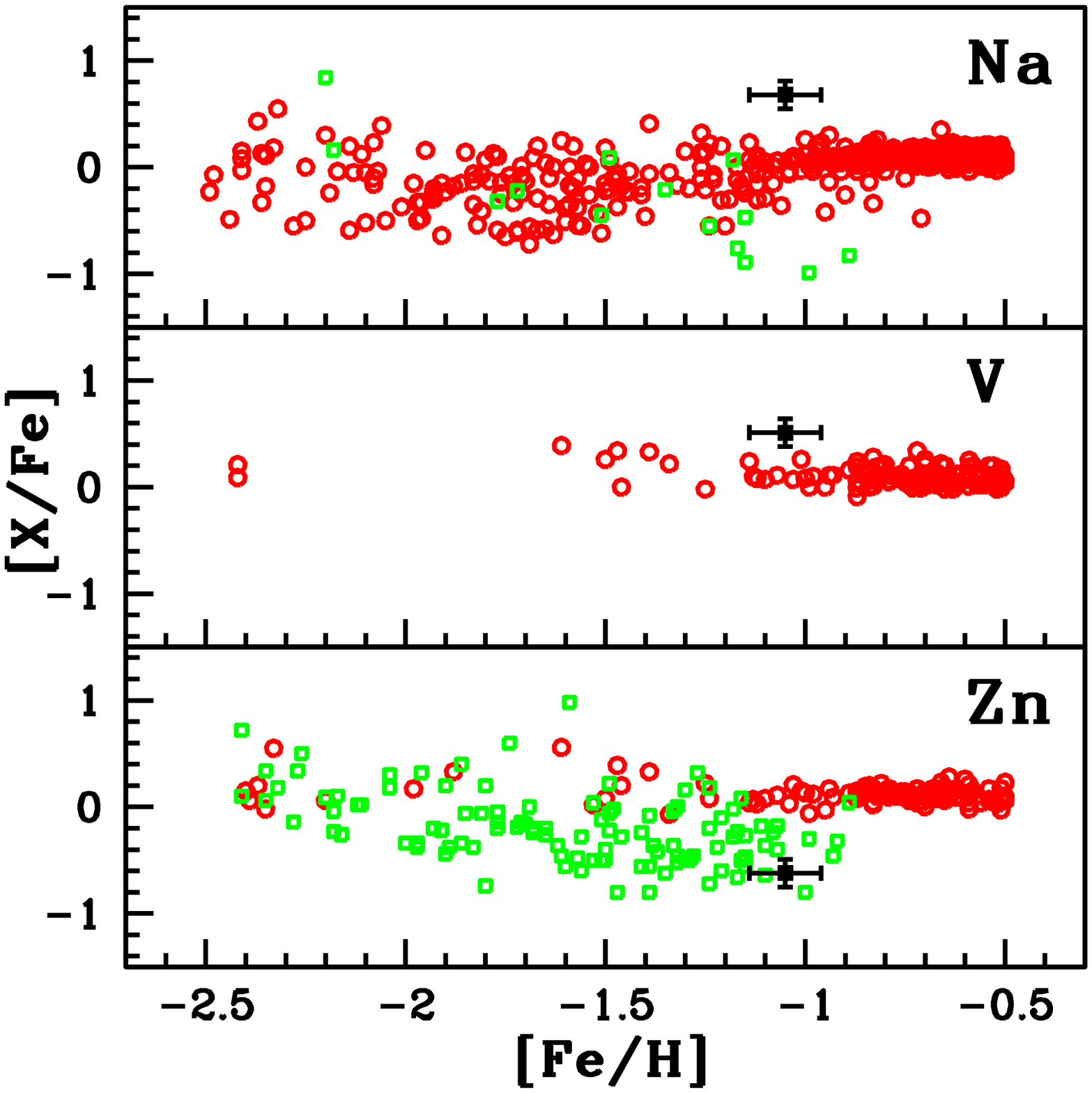}
\caption{Observed Na, V, and Zn abundance ratios of J045 (filled square) with respect to metallicity [Fe/H].   
Red open circles are used to represent the  Galactic normal giants from literature 
\citep{Honda_2004, Venn_2004, Aoki_2005, Aoki_2007, Reddy_2006, Luck_2007, Hansen_2016c, Yoon_2016}.
Green open squares represent Sculptor dwarf galaxy stars from literature \citep{Skuladottir_2017, Hill_2019}.
} \label{Na_V_Zn}
\end{figure}

\begin{figure}
\centering
\includegraphics[width=\columnwidth]{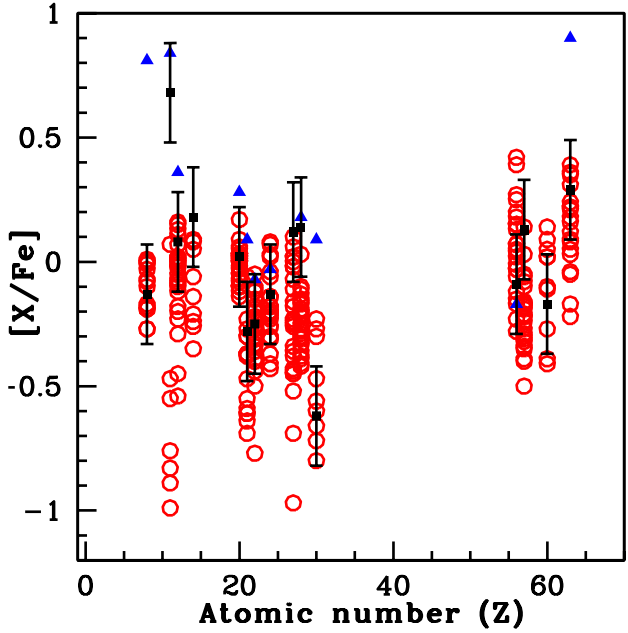}
\caption{Comparison of observed elemental abundances of J045 (filled square) and the sculptor
dwarf galaxy stars (red open circles) from literature \citep{Skuladottir_2017, Hill_2019}.
The elemental abundance ratios of the star UET0130 with [Na/Fe]$\sim$0.84 
is shown in the plot with blue filled triangle.
} \label{sculptor_UET}
\end{figure}

 Na is also enhanced with [Na/Fe] $\sim$ 0.68. Such high abundance values for Na are generally not found to be reported for normal metal-poor Galactic stars of similar metallicity (Figure \ref{Na_V_Zn}). Among the Dwarf Sculptor Galaxy stars one object UET0130 with a metallicity of $-$2.2 is known to show very high Na abundance with [Na/Fe] = 0.84 (Figure \ref{sculptor_UET}).

\par Among the Fe-peak elements, the Mn abundance is found to be near-solar, whereas, Co and Ni are found to be marginally enhanced . Sc and Ti are found to be  under-abundant  with [Sc, Ti/Fe] ${\sim}$ $-$ 0.28 and $-$0.25 respectively.  As seen from Figure \ref{light comparison} (upper panel)  these values match closely with Sculptor Dwarf Galaxy stars. 

\par The  [$\alpha$/Fe] ratios observed in J045 
are generally noticed  in dwarf spheroidal galaxies at closer metallicity  \citep{Koch_2008, Kirby_2009}. The observed [Sc/Fe] ${\sim}$ $-$0.28 and  [Ti/Fe] ${\sim}$ $-$0.25 in J045 are also seen as typical values in many Sculptor Dwarf Galaxy stars \citep{Skuladottir_2017, Hill_2019}.

\par The abundance of V with [V/Fe] = 0.51  is higher than generally noticed in metal-poor Galactic stars. In a recent work,  \cite{Ou_2020} have derived abundance of Vanadium from a sample of 255 metal-poor stars in the metallicity range
$-$4.0 ${\le}$ [Fe/H] ${\le}$ $-$1.0, and found that [V I/Fe] = $-$0.10$\pm$ 0.01 from 128 stars and 
[V II/Fe] = 0.13$\pm$ 0.01 from 220 stars. The authors suspect this offset to be due to departures from 
LTE and recommend using [V II/Fe II]  values which is enhanced relative to 
solar ratio. We could not compare our estimate of vanadium abundance  with  Sculptor Dwarf Galaxy stars  as  literature values are not available.     

\par The derived abundance of Zn  with [Zn/Fe] = $-$0.62 is found to be very different from its counterparts observed
  in Galactic giants and halo stars of similar metallicity. If we take account of the Non-LTE correction
  ($<$ 0.1 dex) zinc abundance
  will slightly enhance but still remains distinctly different from the Galactic trend.
  \cite{Saito_2009} studied  the trend of [Zn/Fe] with respect to metallicity
  in the range $-$4.2 $\leq$ [Fe/H] $\leq$ +0.5, considering a large sample of stars (434 stars)
   belonging to the population of 
  Galactic thin and thick disk as well as  halo, and confirmed a trend where  [Zn/Fe] decreases with 
  increase of [Fe/H]. [Zn/Fe] was found to follow a nearly flat trend  in the range of metallicity
   $-$2.0 $\leq$ [Fe/H] $\leq$ +0.5, with a slight enhancement of $\sim$ +0.007 dex
  upto [Fe/H ] $<$ $-$1.0. The enhancement was found to be more pronounced in the range
   $-$1.0 $\leq$ [Fe/H] $\leq$ $-$0.5, and the trend  is found to gradually decrease
  to solar value in [Fe/H] $>$ $-$0.5.  This observed trend of [Zn/Fe] in the Galactic stars is  very different
  from those observed in Sculptor Dwarf Galaxy (dSph)  stars. As shown  by \cite{Skuladottir_2017},
 the Sculptor Dwarf galaxy stars show a  decreasing trend of [Zn/Fe] ratios at [Fe/H] $>$ $-$1.8 with a
 large star-to-star scatter  ranging from  $-$0.8 to $<$ 0.4. A comparison of the Zn abundance
of J045 with the counterparts in dSph stars with
similar metallicities (Figure \ref{Na_V_Zn})   have shown
a reasonably good match indicating that J045 could be a possible Sculptor Dwarf Galaxy escapee.   

\par According to \cite{Skuladottir_2017}, the [Zn/Fe] trend in the metallicity range $-$2.5 $<$ [Fe/H] $<$ $-$1.0   observed in the Sculptor Dwarf galaxy stars can be attributed to SNe Ia that causes the slope of [Zn/Fe] to be steeper than that in a lower metallicity. Since SNe Ia is not able to produce Zn, stars formed out of a medium contaminated by the ejecta of SNe Ia are expected to show low [Zn/Fe] values and in particular, the  
stars with [Zn/Fe] $<$ $-$0.5 are believed to be formed from a medium which is predominantly influenced by the ejecta of SNe Ia.

\par 
Based on chemodynamical simulations of dwarf galaxies, \cite{Hirai_2018}  suggested that the electron-capture supernovae (ECSNe) can be one of the sources that can produce Zn as they could not 
justify  the 
observed trends if ECSNe is not considered as a source of Zn.

\par
\begin{figure}
\centering
\includegraphics[width=\columnwidth]{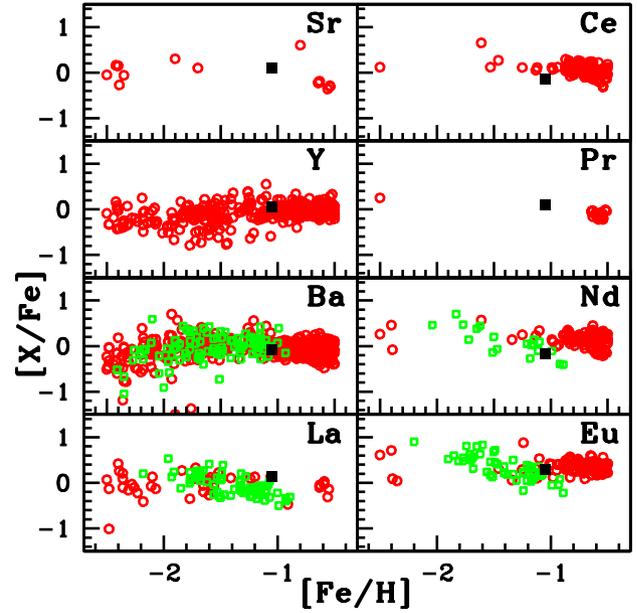}
\caption{Same as Figure \ref{light comparison}, but for heavy elements.} \label{heavy comparison}
\end{figure}

 The abundance of neutron-capture elements Sr, Y, Ba, Pr, Sm are near-solar and La is marginally enhanced
in J045. The abundance of  Ce and Nd are marginally under-abundant, whereas Eu is mildly enhanced. 
In figures \ref{light comparison} and \ref{heavy comparison} we have compared  the estimated abundance ratios of light and heavy elements in J045  with the corresponding ratios  measured in normal Galactic giants  and Sculptor dwarf galaxy stars. The observed abundance pattern in J045 is found to  match well with that of Sculptor dwarf galaxy stars.

We have examined if the accretion of material from a binary companion could be a possible explanation for the non-typical abundance pattern observed in J045. The CH, CEMP-s, and CEMP-r/s stars are metal-poor stars that are believed to have been received  gas transferred from an AGB companion. However, with an estimated [C/Fe] ${\sim}$ 0.11,
the star is neither carbon-enhanced nor s-process enhanced with [X/Fe] ${\sim}$ solar values, where X stands for heavy elements.

\subsection{Mixing diagnostic}

\begin{figure}
\centering
\includegraphics[width=\columnwidth]{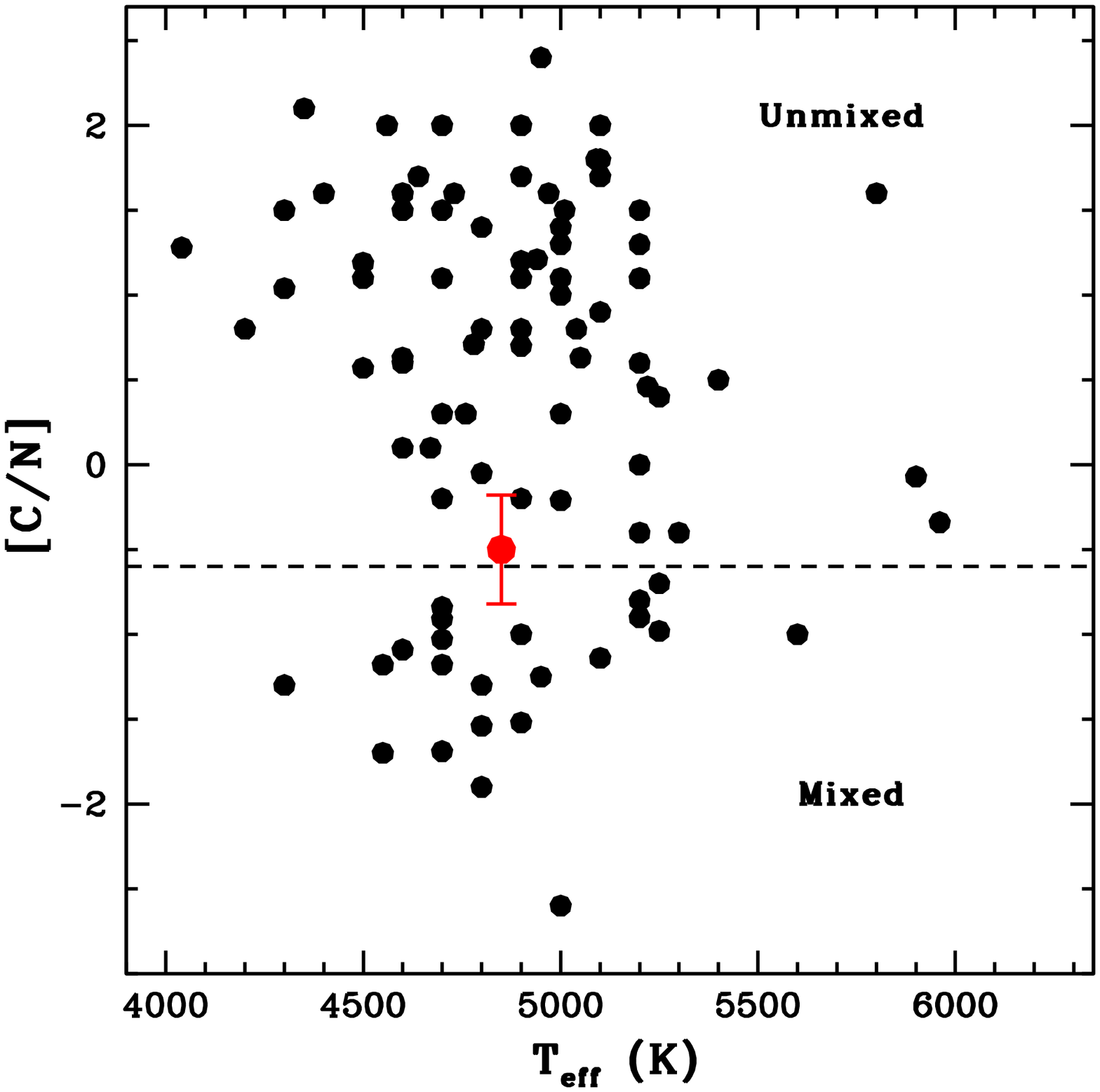}
\caption{Position of J045 (Filled red circle) in the [C/N] vs. T$\rm_{eff}$ diagram. Filled black circles 
indicate the CEMP stars and the normal giants from literature \citep{Spite_2006, Aoki_2007, Luck_2007, Hansen_2014, Goswami_2016, Purandardas_2019a, Purandardas_2019b, Goswami_2021, Purandardas_2021a, Purandardas_2021b, Shejeelammal_2021a, Shejeelammal_2021b}.  The dashed line at [C/N] = $-$0.6 separates the two groups "Mixed" and "Unmixed". The location of J045 in this plot shows it to be "Unmixed", however, when the lower limit of error is considered the object falls into the "Mixed" group.}\label{cn mixing}
\end{figure}

\begin{figure}
\centering
\includegraphics[width=\columnwidth]{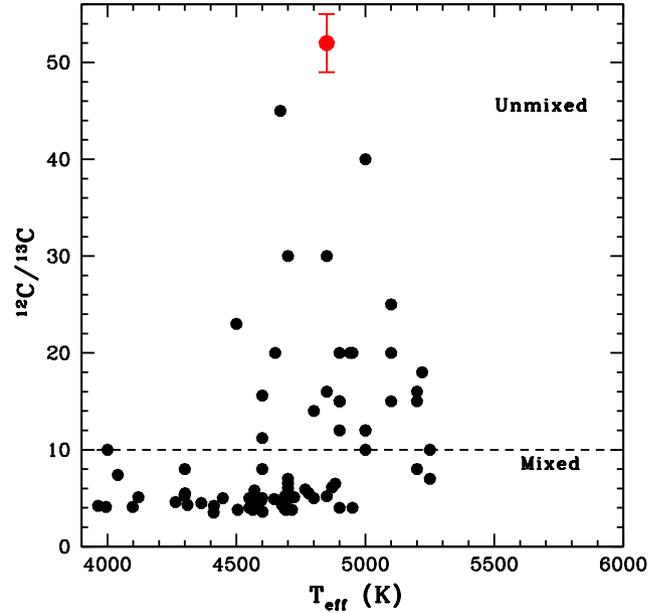}
\caption{Position of J045 (Filled red circle) in the $^{12}$C/$^{13}$C vs. T$\rm_{eff}$ diagram. Filled black circles indicate the CEMP stars and normal giants from literature \citep{Spite_2006, Aoki_2007, Maas_2019}. The dashed line at $^{12}$C/$^{13}$C = 10 separates the groups "Mixed" and "Unmixed".}  \label{cir mixing}
\end{figure}

 Several mixing processes are found to occur in stars in their giant phase of evolution \citep{Spite_2006, Gratton_2000}. Since the object J045 is found to be a giant
 (Figure \ref{track}, upper panel), it is important to examine  whether any internal mixing processes have altered the surface chemical composition of the star. The [C/N]  and $^{12}$C/$^{13}$C ratios can give important clues regarding the occurrence of mixing.  \cite{Spite_2005} have shown that stars with [C/N] $>$ $-$0.60 and $^{12}$C/$^{13}$C $>$  10 are unmixed stars.
   We have therefore checked for the signatures of any internal mixing based on the estimated [C/N] ratio and $^{12}$C/$^{13}$C ratios (Figures \ref{cn mixing} and \ref{cir mixing}). Following the  criteria 
of \cite{Spite_2005},  with [C/N] = $-$0.5 and $^{12}$C/$^{13}$C = 52 the star J045 does not seem to have undergone any internal mixing. 
When we consider the lower limit of the error ($-$0.32)  in [C/N] ratio, the program star is well mixed.  But considering the upper limit of the error in [C/N] ratio, the program star remains unmixed.  
The estimated value of carbon isotopic ratio shows that the programme star is unmixed even within the error limits.
   The estimated high  carbon isotopic ratio in J045 ($\sim$ 52) is   not unusual. For example, \cite{Lucatello_2003, Aoki_2007, Purandardas_2019b} had reported $^{12}$C/$^{13}$C in the range 40 - 60 for giants of similar metallicity.  

\section{Conclusion} \label{section conclusion}
This work presents the results of a detailed  spectroscopic analysis of the object J045 using high
resolution spectra.
The program star is presented in the list of possible carbon stars of LAMOST DR2 catalog. 
However, from our analysis we show that the program star does not show any spectral properties of 
a typical carbon star.
 From our analysis, we find that J045 is a metal-poor giant of metallicity [Fe/H] $-$1.05  with very unusual elemental abundances. The observed abundance pattern is not compatible with abundances characteristic of Galactic metal-poor stars of similar metallicity, but,  matches  quite closely with those observed in Sculptor Dwarf galaxy stars. This observational evidences prompted us to argue that J045 is likely a Sculptor Dwarf Galaxy escapee.

\par While the carbon abundance is estimated to be near-solar, N is over abundant with [N/Fe] $\sim$ 0.61. Unlike in Galactic metal-poor stars, oxygen is found to be marginally under-abundant with [O/Fe] $\sim$ $-$0.13
and $\rm<[\alpha/Fe]>$ $\sim$ 0.09. Na is found to be slightly enhanced with [Na/Fe] $\sim$ 0.68.   Sc and Ti are found to be under-abundant.  While the Fe-peak element Mn is determined to be near-solar,  Ni and Co are estimated to be marginally enhanced.   Among the neutron-capture elements, Eu is marginally enhanced with [Eu/Fe] $\sim$ 0.29.  Sr, Y, Ba, Pr, and Sm show near-solar values. While  La is marginally enhanced, 
Ce and Nd are found to be marginally under-abundant. With the estimated [Ba/Eu] $\sim$ $-$ 0.38, the star falls into the category of neutron-capture-rich star of r-I type  of \cite{Beers_2005}. 

\par The estimated luminosity is high with log(L/L$_{\odot}$)   ${\sim}$ 2.27. Such high values would indicate  that the star might have undergone internal mixing. However, the estimates of [C/N] and $^{12}$C/$^{13}$C ratios show that the object has not undergone any significant internal mixing. 
We however note that, if we consider the lower limit of error in [C/N] ratio, the star is well mixed.
From the kinematic analysis, we find that the program star is a thin disk object.
\appendix

\section*{Acknowledgements}
We thank the referee Prof. Chris Sneden for his constructive and  insightful comments that have significantly improved the manuscript.
The facilities at IAO and CREST are operated by the Indian Institute of Astrophysics, Bangalore.  We thank the staff members of IAO and CREST for help with observations. The DST SERB project EMR/2016/005283 is gratefully acknowledged.  MP was a junior research fellow under this project.
AG, DM, and SJ are  grateful for the support received from  the Indo-Thailand Programme of co-operation in Science and Technology through the Indo-Thai joint project DST/INT/Thai/P-16/2019. This work made use of the SIMBAD astronomical database, operated at CDS, Strasbourg, France, the NASA ADS, USA, and data from the European Space Agency (ESA) mission Gaia (https://www.cosmos.esa.int/gaia), processed by the Gaia Data Processing and Analysis Consortium (DPAC), (https://www.cosmos.esa.int/web/gaia/dpac/consortium). Based on data collected using HCT/HESP. Funding for the TESS mission is provided by NASA's Science Mission directorate.\\

\begin{center}
{\bf Data Availability}\\
\end{center}
\noindent The data underlying this article will be shared on reasonable request to the corresponding author.

\bibliographystyle{mnras}
\bibliography{Bibliography}

\begin{thebibliography}{}
\makeatletter
\relax
\def\mn@urlcharsother{\let\do\@makeother \do\$\do\&\do\#\do\^\do\_\do\%\do\~}
\def\mn@doi{\begingroup\mn@urlcharsother \@ifnextchar [ {\mn@doi@}
  {\mn@doi@[]}}
\def\mn@doi@[#1]#2{\def\@tempa{#1}\ifx\@tempa\@empty \href
  {http://dx.doi.org/#2} {doi:#2}\else \href {http://dx.doi.org/#2} {#1}\fi
  \endgroup}
\def\mn@eprint#1#2{\mn@eprint@#1:#2::\@nil}
\def\mn@eprint@arXiv#1{\href {http://arxiv.org/abs/#1} {{\tt arXiv:#1}}}
\def\mn@eprint@dblp#1{\href {http://dblp.uni-trier.de/rec/bibtex/#1.xml}
  {dblp:#1}}
\def\mn@eprint@#1:#2:#3:#4\@nil{\def\@tempa {#1}\def\@tempb {#2}\def\@tempc
  {#3}\ifx \@tempc \@empty \let \@tempc \@tempb \let \@tempb \@tempa \fi \ifx
  \@tempb \@empty \def\@tempb {arXiv}\fi \@ifundefined
  {mn@eprint@\@tempb}{\@tempb:\@tempc}{\expandafter \expandafter \csname
  mn@eprint@\@tempb\endcsname \expandafter{\@tempc}}}

\bibitem[\protect\citeauthoryear{{Alksnis}, {Balklavs}, {Dzervitis}, {Eglitis},
  {Paupers}  \& {Pundure}}{{Alksnis} et~al.}{2001}]{Alksnis_2001}
{Alksnis} A.,  {Balklavs} A.,  {Dzervitis} U.,  {Eglitis} I.,  {Paupers} O.,
  {Pundure} I.,  2001, \mn@doi [Baltic Astronomy] {10.1515/astro-2001-1-202},
  \href {https://ui.adsabs.harvard.edu/abs/2001BaltA..10....1A} {10, 1}

\bibitem[\protect\citeauthoryear{{Allende Prieto}, {Lambert}  \&
  {Asplund}}{{Allende Prieto} et~al.}{2001}]{Prieto_2001}
{Allende Prieto} C.,  {Lambert} D.~L.,   {Asplund} M.,  2001, \mn@doi [\apjl]
  {10.1086/322874}, \href
  {https://ui.adsabs.harvard.edu/abs/2001ApJ...556L..63A} {556, L63}

\bibitem[\protect\citeauthoryear{{Alonso}, {Arribas}  \&
  {Mart{\'\i}nez-Roger}}{{Alonso} et~al.}{1999}]{Alonso_1999}
{Alonso} A.,  {Arribas} S.,   {Mart{\'\i}nez-Roger} C.,  1999, \mn@doi [\aaps]
  {10.1051/aas:1999521}, \href
  {https://ui.adsabs.harvard.edu/abs/1999A&AS..140..261A} {140, 261}

\bibitem[\protect\citeauthoryear{{Aoki} et~al.,}{{Aoki}
  et~al.}{2005}]{Aoki_2005}
{Aoki} W.,  et~al., 2005, \mn@doi [\apj] {10.1086/432862}, \href
  {https://ui.adsabs.harvard.edu/abs/2005ApJ...632..611A} {632, 611}

\bibitem[\protect\citeauthoryear{{Aoki}, {Beers}, {Christlieb}, {Norris},
  {Ryan}  \& {Tsangarides}}{{Aoki} et~al.}{2007}]{Aoki_2007}
{Aoki} W.,  {Beers} T.~C.,  {Christlieb} N.,  {Norris} J.~E.,  {Ryan} S.~G.,
  {Tsangarides} S.,  2007, \mn@doi [\apj] {10.1086/509817}, \href
  {https://ui.adsabs.harvard.edu/abs/2007ApJ...655..492A} {655, 492}

\bibitem[\protect\citeauthoryear{{Asplund}, {Grevesse}, {Sauval}  \&
  {Scott}}{{Asplund} et~al.}{2009}]{Asplund_2009}
{Asplund} M.,  {Grevesse} N.,  {Sauval} A.~J.,   {Scott} P.,  2009, \mn@doi
  [\araa] {10.1146/annurev.astro.46.060407.145222}, \href
  {https://ui.adsabs.harvard.edu/abs/2009ARA&A..47..481A} {47, 481}

\bibitem[\protect\citeauthoryear{{Bai} et~al.,}{{Bai} et~al.}{2016}]{Bai_2016}
{Bai} Y.,  et~al., 2016, \mn@doi [Research in Astronomy and Astrophysics]
  {10.1088/1674-4527/16/7/107}, \href
  {https://ui.adsabs.harvard.edu/abs/2016RAA....16..107B} {16, 107}

\bibitem[\protect\citeauthoryear{{Beers} \& {Christlieb}}{{Beers} \&
  {Christlieb}}{2005}]{Beers_2005}
{Beers} T.~C.,  {Christlieb} N.,  2005, \mn@doi [\araa]
  {10.1146/annurev.astro.42.053102.134057}, \href
  {https://ui.adsabs.harvard.edu/abs/2005ARA&A..43..531B} {43, 531}

\bibitem[\protect\citeauthoryear{{Bensby}, {Feltzing}  \&
  {Lundstr{\"o}m}}{{Bensby} et~al.}{2003}]{Bensby_2003}
{Bensby} T.,  {Feltzing} S.,   {Lundstr{\"o}m} I.,  2003, \mn@doi [\aap]
  {10.1051/0004-6361:20031213}, \href
  {https://ui.adsabs.harvard.edu/abs/2003A&A...410..527B} {410, 527}

\bibitem[\protect\citeauthoryear{{Bensby}, {Feltzing}  \&
  {Lundstr{\"o}m}}{{Bensby} et~al.}{2004}]{Bensby_2004}
{Bensby} T.,  {Feltzing} S.,   {Lundstr{\"o}m} I.,  2004, \mn@doi [\aap]
  {10.1051/0004-6361:20031655}, \href
  {https://ui.adsabs.harvard.edu/abs/2004A&A...415..155B} {415, 155}

\bibitem[\protect\citeauthoryear{{Bergemann}, {Hansen}, {Bautista}  \&
  {Ruchti}}{{Bergemann} et~al.}{2012}]{Bergemann_2012}
{Bergemann} M.,  {Hansen} C.~J.,  {Bautista} M.,   {Ruchti} G.,  2012, \mn@doi
  [\aap] {10.1051/0004-6361/201219406}, \href
  {https://ui.adsabs.harvard.edu/abs/2012A&A...546A..90B} {546, A90}

\bibitem[\protect\citeauthoryear{{Brasseur}, {Phillip}, {Fleming}, {Mullally}
  \& {White}}{{Brasseur} et~al.}{2019}]{Brasseur_2019}
{Brasseur} C.~E.,  {Phillip} C.,  {Fleming} S.~W.,  {Mullally} S.~E.,   {White}
  R.~L.,  2019, {Astrocut: Tools for creating cutouts of TESS images}
  (\mn@eprint {ascl} {1905.007})

\bibitem[\protect\citeauthoryear{{Chen}, {Vergely}, {Valette}  \&
  {Carraro}}{{Chen} et~al.}{1998}]{Chen_1998}
{Chen} B.,  {Vergely} J.~L.,  {Valette} B.,   {Carraro} G.,  1998, \aap, \href
  {https://ui.adsabs.harvard.edu/abs/1998A&A...336..137C} {336, 137}

\bibitem[\protect\citeauthoryear{{Chen}, {Nissen}  \& {Zhao}}{{Chen}
  et~al.}{2004}]{Chen_2004}
{Chen} Y.~Q.,  {Nissen} P.~E.,   {Zhao} G.,  2004, \mn@doi [\aap]
  {10.1051/0004-6361:20047191}, \href
  {https://ui.adsabs.harvard.edu/abs/2004A&A...425..697C} {425, 697}

\bibitem[\protect\citeauthoryear{{Christlieb}, {Green}, {Wisotzki}  \&
  {Reimers}}{{Christlieb} et~al.}{2001}]{Christlieb_2001a}
{Christlieb} N.,  {Green} P.~J.,  {Wisotzki} L.,   {Reimers} D.,  2001, \mn@doi
  [\aap] {10.1051/0004-6361:20010814}, \href
  {https://ui.adsabs.harvard.edu/abs/2001A&A...375..366C} {375, 366}

\bibitem[\protect\citeauthoryear{{Cutri} et~al.,}{{Cutri}
  et~al.}{2003}]{Cutri_2003}
{Cutri} R.~M.,  et~al., 2003, VizieR Online Data Catalog, \href
  {https://ui.adsabs.harvard.edu/abs/2003yCat.2246....0C} {p. II/246}

\bibitem[\protect\citeauthoryear{{Fern{\'a}ndez-Trincado}
  et~al.,}{{Fern{\'a}ndez-Trincado} et~al.}{2019}]{Fernandez_2019}
{Fern{\'a}ndez-Trincado} J.~G.,  et~al., 2019, \mn@doi [\aap]
  {10.1051/0004-6361/201935369}, \href
  {https://ui.adsabs.harvard.edu/abs/2019A&A...631A..97F} {631, A97}

\bibitem[\protect\citeauthoryear{{Gaia Collaboration} et~al.,}{{Gaia
  Collaboration} et~al.}{2016}]{Gaia_2016}
{Gaia Collaboration} et~al., 2016, \mn@doi [\aap]
  {10.1051/0004-6361/201629272}, \href
  {https://ui.adsabs.harvard.edu/abs/2016A&A...595A...1G} {595, A1}

\bibitem[\protect\citeauthoryear{{Gaia Collaboration} et~al.,}{{Gaia
  Collaboration} et~al.}{2018}]{Gaia_2018}
{Gaia Collaboration} et~al., 2018, \mn@doi [\aap]
  {10.1051/0004-6361/201832865}, \href
  {https://ui.adsabs.harvard.edu/abs/2018A&A...616A..11G} {616, A11}

\bibitem[\protect\citeauthoryear{{Gigoyan}, {Hambaryan}  \&
  {Azzopardi}}{{Gigoyan} et~al.}{1998}]{Gigoyan_1998}
{Gigoyan} K.~S.,  {Hambaryan} V.~V.,   {Azzopardi} M.,  1998, \mn@doi
  [Astrophysics] {10.1007/BF02894661}, \href
  {https://ui.adsabs.harvard.edu/abs/1998Ap.....41..356G} {41, 356}

\bibitem[\protect\citeauthoryear{{Girardi}, {Bressan}, {Bertelli}  \&
  {Chiosi}}{{Girardi} et~al.}{2000}]{Girardi_2000}
{Girardi} L.,  {Bressan} A.,  {Bertelli} G.,   {Chiosi} C.,  2000, \mn@doi
  [\aaps] {10.1051/aas:2000126}, \href
  {https://ui.adsabs.harvard.edu/abs/2000A&AS..141..371G} {141, 371}

\bibitem[\protect\citeauthoryear{{Goswami}}{{Goswami}}{2005}]{Goswami_2005}
{Goswami} A.,  2005, \mn@doi [\mnras] {10.1111/j.1365-2966.2005.08917.x}, \href
  {https://ui.adsabs.harvard.edu/abs/2005MNRAS.359..531G} {359, 531}

\bibitem[\protect\citeauthoryear{{Goswami}, {Aoki}, {Beers}, {Christlieb},
  {Norris}, {Ryan}  \& {Tsangarides}}{{Goswami} et~al.}{2006}]{Goswami_2006}
{Goswami} A.,  {Aoki} W.,  {Beers} T.~C.,  {Christlieb} N.,  {Norris} J.~E.,
  {Ryan} S.~G.,   {Tsangarides} S.,  2006, \mn@doi [\mnras]
  {10.1111/j.1365-2966.2006.10877.x}, \href
  {https://ui.adsabs.harvard.edu/abs/2006MNRAS.372..343G} {372, 343}

\bibitem[\protect\citeauthoryear{{Goswami}, {Bama}, {Shantikumar}  \&
  {Devassy}}{{Goswami} et~al.}{2007}]{Goswami_2007}
{Goswami} A.,  {Bama} P.,  {Shantikumar} N.~S.,   {Devassy} D.,  2007, Bulletin
  of the Astronomical Society of India, \href
  {https://ui.adsabs.harvard.edu/abs/2007BASI...35..339G} {35, 339}

\bibitem[\protect\citeauthoryear{{Goswami}, {Karinkuzhi}  \&
  {Shantikumar}}{{Goswami} et~al.}{2010}]{goswami_2010b}
{Goswami} A.,  {Karinkuzhi} D.,   {Shantikumar} N.~S.,  2010, \mn@doi [\mnras]
  {10.1111/j.1365-2966.2009.15939.x}, \href
  {https://ui.adsabs.harvard.edu/abs/2010MNRAS.402.1111G} {402, 1111}

\bibitem[\protect\citeauthoryear{{Goswami}, {Aoki}  \& {Karinkuzhi}}{{Goswami}
  et~al.}{2016}]{Goswami_2016}
{Goswami} A.,  {Aoki} W.,   {Karinkuzhi} D.,  2016, \mn@doi [\mnras]
  {10.1093/mnras/stv2011}, \href
  {https://ui.adsabs.harvard.edu/abs/2016MNRAS.455..402G} {455, 402}

\bibitem[\protect\citeauthoryear{{Goswami}, {Rathour}  \& {Goswami}}{{Goswami}
  et~al.}{2021}]{Goswami_2021}
{Goswami} P.~P.,  {Rathour} R.~S.,   {Goswami} A.,  2021, \mn@doi [\aap]
  {10.1051/0004-6361/202038258}, \href
  {https://ui.adsabs.harvard.edu/abs/2021A&A...649A..49G} {649, A49}

\bibitem[\protect\citeauthoryear{{Gratton}, {Sneden}, {Carretta}  \&
  {Bragaglia}}{{Gratton} et~al.}{2000}]{Gratton_2000}
{Gratton} R.~G.,  {Sneden} C.,  {Carretta} E.,   {Bragaglia} A.,  2000, \aap,
  \href {https://ui.adsabs.harvard.edu/abs/2000A&A...354..169G} {354, 169}

\bibitem[\protect\citeauthoryear{{Hansen} et~al.,}{{Hansen}
  et~al.}{2014}]{Hansen_2014}
{Hansen} T.,  et~al., 2014, \mn@doi [\apj] {10.1088/0004-637X/787/2/162}, \href
  {https://ui.adsabs.harvard.edu/abs/2014ApJ...787..162H} {787, 162}

\bibitem[\protect\citeauthoryear{{Hansen} et~al.,}{{Hansen}
  et~al.}{2016}]{Hansen_2016c}
{Hansen} C.~J.,  et~al., 2016, \mn@doi [\aap] {10.1051/0004-6361/201526895},
  \href {https://ui.adsabs.harvard.edu/abs/2016A&A...588A..37H} {588, A37}

\bibitem[\protect\citeauthoryear{{Hill} et~al.,}{{Hill}
  et~al.}{2019}]{Hill_2019}
{Hill} V.,  et~al., 2019, \mn@doi [\aap] {10.1051/0004-6361/201833950}, \href
  {https://ui.adsabs.harvard.edu/abs/2019A&A...626A..15H} {626, A15}

\bibitem[\protect\citeauthoryear{{Hirai}, {Saitoh}, {Ishimaru}  \&
  {Wanajo}}{{Hirai} et~al.}{2018}]{Hirai_2018}
{Hirai} Y.,  {Saitoh} T.~R.,  {Ishimaru} Y.,   {Wanajo} S.,  2018, \mn@doi
  [\apj] {10.3847/1538-4357/aaaabc}, \href
  {https://ui.adsabs.harvard.edu/abs/2018ApJ...855...63H} {855, 63}

\bibitem[\protect\citeauthoryear{{Honda}, {Aoki}, {Kajino}, {Ando}, {Beers},
  {Izumiura}, {Sadakane}  \& {Takada-Hidai}}{{Honda} et~al.}{2004}]{Honda_2004}
{Honda} S.,  {Aoki} W.,  {Kajino} T.,  {Ando} H.,  {Beers} T.~C.,  {Izumiura}
  H.,  {Sadakane} K.,   {Takada-Hidai} M.,  2004, \mn@doi [\apj]
  {10.1086/383406}, \href
  {https://ui.adsabs.harvard.edu/abs/2004ApJ...607..474H} {607, 474}

\bibitem[\protect\citeauthoryear{{Ibata}, {Lewis}, {Irwin}, {Totten}  \&
  {Quinn}}{{Ibata} et~al.}{2001}]{Ibata_2001}
{Ibata} R.,  {Lewis} G.~F.,  {Irwin} M.,  {Totten} E.,   {Quinn} T.,  2001,
  \mn@doi [\apj] {10.1086/320060}, \href
  {https://ui.adsabs.harvard.edu/abs/2001ApJ...551..294I} {551, 294}

\bibitem[\protect\citeauthoryear{{Ishigaki}, {Chiba}  \& {Aoki}}{{Ishigaki}
  et~al.}{2012}]{Ishigaki_2012}
{Ishigaki} M.~N.,  {Chiba} M.,   {Aoki} W.,  2012, \mn@doi [\apj]
  {10.1088/0004-637X/753/1/64}, \href
  {https://ui.adsabs.harvard.edu/abs/2012ApJ...753...64I} {753, 64}

\bibitem[\protect\citeauthoryear{{Ishigaki}, {Aoki}  \& {Chiba}}{{Ishigaki}
  et~al.}{2013}]{Ishigaki_2013}
{Ishigaki} M.~N.,  {Aoki} W.,   {Chiba} M.,  2013, \mn@doi [\apj]
  {10.1088/0004-637X/771/1/67}, \href
  {https://ui.adsabs.harvard.edu/abs/2013ApJ...771...67I} {771, 67}

\bibitem[\protect\citeauthoryear{{Ji}, {Cui}, {Liu}, {Luo}, {Zhao}  \&
  {Zhang}}{{Ji} et~al.}{2016}]{Ji_2016}
{Ji} W.,  {Cui} W.,  {Liu} C.,  {Luo} A.,  {Zhao} G.,   {Zhang} B.,  2016,
  \mn@doi [\apjs] {10.3847/0067-0049/226/1/1}, \href
  {https://ui.adsabs.harvard.edu/abs/2016ApJS..226....1J} {226, 1}

\bibitem[\protect\citeauthoryear{{Johnson} \& {Soderblom}}{{Johnson} \&
  {Soderblom}}{1987}]{Johnson_1987}
{Johnson} D. R.~H.,  {Soderblom} D.~R.,  1987, \mn@doi [\aj] {10.1086/114370},
  \href {https://ui.adsabs.harvard.edu/abs/1987AJ.....93..864J} {93, 864}

\bibitem[\protect\citeauthoryear{{J{\"o}nsson} et~al.,}{{J{\"o}nsson}
  et~al.}{2020}]{Jonsson_2020}
{J{\"o}nsson} H.,  et~al., 2020, \mn@doi [\aj] {10.3847/1538-3881/aba592},
  \href {https://ui.adsabs.harvard.edu/abs/2020AJ....160..120J} {160, 120}

\bibitem[\protect\citeauthoryear{{Kirby}, {Guhathakurta}, {Bolte}, {Sneden}  \&
  {Geha}}{{Kirby} et~al.}{2009}]{Kirby_2009}
{Kirby} E.~N.,  {Guhathakurta} P.,  {Bolte} M.,  {Sneden} C.,   {Geha} M.~C.,
  2009, \mn@doi [\apj] {10.1088/0004-637X/705/1/328}, \href
  {https://ui.adsabs.harvard.edu/abs/2009ApJ...705..328K} {705, 328}

\bibitem[\protect\citeauthoryear{{Koch} \& {McWilliam}}{{Koch} \&
  {McWilliam}}{2008}]{Koch_2008}
{Koch} A.,  {McWilliam} A.,  2008, \mn@doi [\aj]
  {10.1088/0004-6256/135/4/1551}, \href
  {https://ui.adsabs.harvard.edu/abs/2008AJ....135.1551K} {135, 1551}

\bibitem[\protect\citeauthoryear{{Lennon}, {Dufton}  \& {Crowley}}{{Lennon}
  et~al.}{2003}]{Lennon_2003}
{Lennon} D.~J.,  {Dufton} P.~L.,   {Crowley} C.,  2003, \mn@doi [\aap]
  {10.1051/0004-6361:20021194}, \href
  {https://ui.adsabs.harvard.edu/abs/2003A&A...398..455L} {398, 455}

\bibitem[\protect\citeauthoryear{{Lightkurve Collaboration}
  et~al.,}{{Lightkurve Collaboration} et~al.}{2018}]{2018ascl.soft12013L}
{Lightkurve Collaboration} et~al., 2018, {Lightkurve: Kepler and TESS time
  series analysis in Python}, Astrophysics Source Code Library (\mn@eprint
  {ascl} {1812.013})

\bibitem[\protect\citeauthoryear{{Lomb}}{{Lomb}}{1976}]{Lomb_1976}
{Lomb} N.~R.,  1976, \mn@doi [\apss] {10.1007/BF00648343}, \href
  {https://ui.adsabs.harvard.edu/abs/1976Ap&SS..39..447L} {39, 447}

\bibitem[\protect\citeauthoryear{{Lucatello}, {Gratton}, {Cohen}, {Beers},
  {Christlieb}, {Carretta}  \& {Ram{\'\i}rez}}{{Lucatello}
  et~al.}{2003}]{Lucatello_2003}
{Lucatello} S.,  {Gratton} R.,  {Cohen} J.~G.,  {Beers} T.~C.,  {Christlieb}
  N.,  {Carretta} E.,   {Ram{\'\i}rez} S.,  2003, \mn@doi [\aj]
  {10.1086/345886}, \href
  {https://ui.adsabs.harvard.edu/abs/2003AJ....125..875L} {125, 875}

\bibitem[\protect\citeauthoryear{{Luck} \& {Heiter}}{{Luck} \&
  {Heiter}}{2007}]{Luck_2007}
{Luck} R.~E.,  {Heiter} U.,  2007, \mn@doi [\aj] {10.1086/513194}, \href
  {https://ui.adsabs.harvard.edu/abs/2007AJ....133.2464L} {133, 2464}

\bibitem[\protect\citeauthoryear{{Luo} et~al.,}{{Luo} et~al.}{2015}]{Luo_2015}
{Luo} A.~L.,  et~al., 2015, \mn@doi [Research in Astronomy and Astrophysics]
  {10.1088/1674-4527/15/8/002}, \href
  {https://ui.adsabs.harvard.edu/abs/2015RAA....15.1095L} {15, 1095}

\bibitem[\protect\citeauthoryear{{Maas}, {Gerber}, {Deibel}  \&
  {Pilachowski}}{{Maas} et~al.}{2019}]{Maas_2019}
{Maas} Z.~G.,  {Gerber} J.~M.,  {Deibel} A.,   {Pilachowski} C.~A.,  2019,
  \mn@doi [\apj] {10.3847/1538-4357/ab1eab}, \href
  {https://ui.adsabs.harvard.edu/abs/2019ApJ...878...43M} {878, 43}

\bibitem[\protect\citeauthoryear{{Martell}, {Smolinski}, {Beers}  \&
  {Grebel}}{{Martell} et~al.}{2011}]{Martell_2011}
{Martell} S.~L.,  {Smolinski} J.~P.,  {Beers} T.~C.,   {Grebel} E.~K.,  2011,
  \mn@doi [\aap] {10.1051/0004-6361/201117644}, \href
  {https://ui.adsabs.harvard.edu/abs/2011A&A...534A.136M} {534, A136}

\bibitem[\protect\citeauthoryear{{Mikolaitis} et~al.,}{{Mikolaitis}
  et~al.}{2019}]{Mikolaitis_2019}
{Mikolaitis} {\v{S}}.,  et~al., 2019, \mn@doi [\aap]
  {10.1051/0004-6361/201835004}, \href
  {https://ui.adsabs.harvard.edu/abs/2019A&A...628A..49M} {628, A49}

\bibitem[\protect\citeauthoryear{{Mishenina}, {Soubiran}, {Kovtyukh}  \&
  {Korotin}}{{Mishenina} et~al.}{2004}]{Mishenina_2004}
{Mishenina} T.~V.,  {Soubiran} C.,  {Kovtyukh} V.~V.,   {Korotin} S.~A.,  2004,
  \mn@doi [\aap] {10.1051/0004-6361:20034454}, \href
  {https://ui.adsabs.harvard.edu/abs/2004A&A...418..551M} {418, 551}

\bibitem[\protect\citeauthoryear{{Ou}, {Roederer}, {Sneden}, {Cowan}, {Lawler},
  {Shectman}  \& {Thompson}}{{Ou} et~al.}{2020}]{Ou_2020}
{Ou} X.,  {Roederer} I.~U.,  {Sneden} C.,  {Cowan} J.~J.,  {Lawler} J.~E.,
  {Shectman} S.~A.,   {Thompson} I.~B.,  2020, \mn@doi [\apj]
  {10.3847/1538-4357/abaa50}, \href
  {https://ui.adsabs.harvard.edu/abs/2020ApJ...900..106O} {900, 106}

\bibitem[\protect\citeauthoryear{{Placco}, {Sneden}, {Roederer}, {Lawler}, {Den
  Hartog}, {Hejazi}, {Maas}  \& {Bernath}}{{Placco} et~al.}{2021}]{Placco_2021}
{Placco} V.~M.,  {Sneden} C.,  {Roederer} I.~U.,  {Lawler} J.~E.,  {Den Hartog}
  E.~A.,  {Hejazi} N.,  {Maas} Z.,   {Bernath} P.,  2021, \mn@doi [Research
  Notes of the American Astronomical Society] {10.3847/2515-5172/abf651}, \href
  {https://ui.adsabs.harvard.edu/abs/2021RNAAS...5...92P} {5, 92}

\bibitem[\protect\citeauthoryear{{Purandardas} \& {Goswami}}{{Purandardas} \&
  {Goswami}}{2021a}]{Purandardas_2021b}
{Purandardas} M.,  {Goswami} A.,  2021a, arXiv e-prints, \href
  {https://ui.adsabs.harvard.edu/abs/2021arXiv210806075P} {p. arXiv:2108.06075}

\bibitem[\protect\citeauthoryear{{Purandardas} \& {Goswami}}{{Purandardas} \&
  {Goswami}}{2021b}]{Purandardas_2021a}
{Purandardas} M.,  {Goswami} A.,  2021b, \mn@doi [\apj]
  {10.3847/1538-4357/abec45}, \href
  {https://ui.adsabs.harvard.edu/abs/2021ApJ...912...74P} {912, 74}

\bibitem[\protect\citeauthoryear{{Purandardas}, {Goswami}  \&
  {Doddamani}}{{Purandardas} et~al.}{2019a}]{Purandardas_2019b}
{Purandardas} M.,  {Goswami} A.,   {Doddamani} V.~H.,  2019a, Bulletin de la
  Societe Royale des Sciences de Liege, \href
  {https://ui.adsabs.harvard.edu/abs/2019BSRSL..88..207P} {88, 207}

\bibitem[\protect\citeauthoryear{{Purandardas}, {Goswami}, {Goswami},
  {Shejeelammal}  \& {Masseron}}{{Purandardas}
  et~al.}{2019b}]{Purandardas_2019a}
{Purandardas} M.,  {Goswami} A.,  {Goswami} P.~P.,  {Shejeelammal} J.,
  {Masseron} T.,  2019b, \mn@doi [\mnras] {10.1093/mnras/stz759}, \href
  {https://ui.adsabs.harvard.edu/abs/2019MNRAS.486.3266P} {486, 3266}

\bibitem[\protect\citeauthoryear{{Reddy}, {Lambert}  \& {Allende
  Prieto}}{{Reddy} et~al.}{2006}]{Reddy_2006}
{Reddy} B.~E.,  {Lambert} D.~L.,   {Allende Prieto} C.,  2006, \mn@doi [\mnras]
  {10.1111/j.1365-2966.2006.10148.x}, \href
  {https://ui.adsabs.harvard.edu/abs/2006MNRAS.367.1329R} {367, 1329}

\bibitem[\protect\citeauthoryear{{Saito}, {Takada-Hidai}, {Honda}  \&
  {Takeda}}{{Saito} et~al.}{2009}]{Saito_2009}
{Saito} Y.-J.,  {Takada-Hidai} M.,  {Honda} S.,   {Takeda} Y.,  2009, \mn@doi
  [\pasj] {10.1093/pasj/61.3.549}, \href
  {https://ui.adsabs.harvard.edu/abs/2009PASJ...61..549S} {61, 549}

\bibitem[\protect\citeauthoryear{{Scargle}}{{Scargle}}{1982}]{Scargle_1982}
{Scargle} J.~D.,  1982, \mn@doi [\apj] {10.1086/160554}, \href
  {https://ui.adsabs.harvard.edu/abs/1982ApJ...263..835S} {263, 835}

\bibitem[\protect\citeauthoryear{{Shejeelammal} \& {Goswami}}{{Shejeelammal} \&
  {Goswami}}{2021}]{Shejeelammal_2021b}
{Shejeelammal} J.,  {Goswami} A.,  2021, \mn@doi [\apj]
  {10.3847/1538-4357/ac1ac9}, \href
  {https://ui.adsabs.harvard.edu/abs/2021ApJ...921...77S} {921, 77}

\bibitem[\protect\citeauthoryear{{Shejeelammal}, {Goswami}  \&
  {Shi}}{{Shejeelammal} et~al.}{2021}]{Shejeelammal_2021a}
{Shejeelammal} J.,  {Goswami} A.,   {Shi} J.,  2021, \mn@doi [\mnras]
  {10.1093/mnras/staa3892}, \href
  {https://ui.adsabs.harvard.edu/abs/2021MNRAS.502.1008S} {502, 1008}

\bibitem[\protect\citeauthoryear{{Sk{\'u}lad{\'o}ttir}, {Tolstoy}, {Salvadori},
  {Hill}  \& {Pettini}}{{Sk{\'u}lad{\'o}ttir} et~al.}{2017}]{Skuladottir_2017}
{Sk{\'u}lad{\'o}ttir} {\'A}.,  {Tolstoy} E.,  {Salvadori} S.,  {Hill} V.,
  {Pettini} M.,  2017, \mn@doi [\aap] {10.1051/0004-6361/201731158}, \href
  {https://ui.adsabs.harvard.edu/abs/2017A&A...606A..71S} {606, A71}

\bibitem[\protect\citeauthoryear{{Sneden}}{{Sneden}}{1973}]{Sneden_1973}
{Sneden} C.~A.,  1973, PhD thesis, THE UNIVERSITY OF TEXAS AT AUSTIN.

\bibitem[\protect\citeauthoryear{{Spite} et~al.,}{{Spite}
  et~al.}{2005}]{Spite_2005}
{Spite} M.,  et~al., 2005, \mn@doi [\aap] {10.1051/0004-6361:20041274}, \href
  {https://ui.adsabs.harvard.edu/abs/2005A&A...430..655S} {430, 655}

\bibitem[\protect\citeauthoryear{{Spite} et~al.,}{{Spite}
  et~al.}{2006}]{Spite_2006}
{Spite} M.,  et~al., 2006, \mn@doi [\aap] {10.1051/0004-6361:20065209}, \href
  {https://ui.adsabs.harvard.edu/abs/2006A&A...455..291S} {455, 291}

\bibitem[\protect\citeauthoryear{{Takeda}, {Hashimoto}, {Taguchi}, {Yoshioka},
  {Takada-Hidai}, {Saito}  \& {Honda}}{{Takeda} et~al.}{2005}]{Takeda_2005}
{Takeda} Y.,  {Hashimoto} O.,  {Taguchi} H.,  {Yoshioka} K.,  {Takada-Hidai}
  M.,  {Saito} Y.,   {Honda} S.,  2005, \mn@doi [\pasj]
  {10.1093/pasj/57.5.751}, \href
  {https://ui.adsabs.harvard.edu/abs/2005PASJ...57..751T} {57, 751}

\bibitem[\protect\citeauthoryear{{Totten} \& {Irwin}}{{Totten} \&
  {Irwin}}{1998}]{Totten_1998}
{Totten} E.~J.,  {Irwin} M.~J.,  1998, \mn@doi [\mnras]
  {10.1046/j.1365-8711.1998.01086.x}, \href
  {https://ui.adsabs.harvard.edu/abs/1998MNRAS.294....1T} {294, 1}

\bibitem[\protect\citeauthoryear{{Venn}, {Irwin}, {Shetrone}, {Tout}, {Hill}
  \& {Tolstoy}}{{Venn} et~al.}{2004}]{Venn_2004}
{Venn} K.~A.,  {Irwin} M.,  {Shetrone} M.~D.,  {Tout} C.~A.,  {Hill} V.,
  {Tolstoy} E.,  2004, \mn@doi [\aj] {10.1086/422734}, \href
  {https://ui.adsabs.harvard.edu/abs/2004AJ....128.1177V} {128, 1177}

\bibitem[\protect\citeauthoryear{{Wu} et~al.,}{{Wu} et~al.}{2011}]{Wu_2011}
{Wu} Y.,  et~al., 2011, \mn@doi [Research in Astronomy and Astrophysics]
  {10.1088/1674-4527/11/8/006}, \href
  {https://ui.adsabs.harvard.edu/abs/2011RAA....11..924W} {11, 924}

\bibitem[\protect\citeauthoryear{{Xiang} et~al.,}{{Xiang}
  et~al.}{2019}]{Xiang_2019}
{Xiang} M.,  et~al., 2019, \mn@doi [\apjs] {10.3847/1538-4365/ab5364}, \href
  {https://ui.adsabs.harvard.edu/abs/2019ApJS..245...34X} {245, 34}

\bibitem[\protect\citeauthoryear{{Yang} et~al.,}{{Yang}
  et~al.}{2016}]{Yang_2016}
{Yang} G.-C.,  et~al., 2016, \mn@doi [Research in Astronomy and Astrophysics]
  {10.1088/1674-4527/16/1/019}, \href
  {https://ui.adsabs.harvard.edu/abs/2016RAA....16...19Y} {16, 19}

\bibitem[\protect\citeauthoryear{{Yoon} et~al.,}{{Yoon}
  et~al.}{2016}]{Yoon_2016}
{Yoon} J.,  et~al., 2016, \mn@doi [\apj] {10.3847/0004-637X/833/1/20}, \href
  {https://ui.adsabs.harvard.edu/abs/2016ApJ...833...20Y} {833, 20}

\makeatother
\end{thebibliography}

\appendix
\section{\bf Line lists }
Lines used for deriving elemental abundances. 

\begin{table*}
\centering
{\bf Table A1 :Equivalent widths (in m\r{A}) of Fe lines used for deriving atmospheric parameters.}
\resizebox{0.68\textwidth}{!}{\begin{tabular}{ccccc}
\hline                       
Wavelength(\r{A}) & Element & $E_{low}$(eV) & log gf & LAMOST J045019.27+394758.7 \\ 
\hline 
4438.350	&	Fe I	&	3.880	&	-1.630	&	26.7(6.46)	\\
4446.830	&		&	3.690	&	-1.330	&	42.7(6.37)	\\
4566.510	&		&	3.300	&	-2.250	&	30.9(6.53)	\\
4635.850	&		&	2.850	&	-2.420	&	46.0(6.55)	\\
4690.140	&		&	3.690	&	-1.640	&	32.6(6.39)	\\
4768.320	&		&	3.690	&	-1.109	&	57.8(6.53)	\\
4789.650	&		&	3.550	&	-0.910	&	62.4(6.30)	\\
4907.740	&		&	3.430	&	-1.840	&	35.7(6.36)	\\
4967.890	&		&	4.190	&	-0.622	&	45.9(6.27)	\\
5079.220	&		&	2.200	&	-2.067	&	86.0(6.54)	\\
5109.650	&		&	4.300	&	-0.980	&	36.7(6.50)	\\
5369.960	&		&	4.370	&	0.350	&	73.8(6.24)	\\
5373.700	&		&	4.470	&	-0.860	&	36.7(6.56)	\\
5379.570	&		&	3.690	&	-1.480	&	42.5(6.43)	\\
5543.940	&		&	4.220	&	-1.140	&	37.1(6.55)	\\
5554.880	&		&	4.580	&	-0.440	&	49.0(6.57)	\\
5618.630	&		&	4.210	&	-1.270	&	33.6(6.58)	\\
5701.550	&		&	2.560	&	-2.216	&	65.1(6.37)	\\
5753.120	&		&	4.260	&	-0.760	&	46.1(6.43)	\\
5859.580	&		&	4.550	&	-0.398	&	41.5(6.28)	\\
5883.810	&		&	3.960	&	-1.360	&	42.4(6.58)	\\
5914.190	&		&	4.610	&	-0.050	&	61.7(6.52)	\\
5956.690	&		&	0.860	&	-4.605	&	53.8(6.35)	\\
6003.020	&		&	3.880	&	-1.120	&	49.3(6.41)	\\
6082.710	&		&	2.220	&	-3.550	&	39.5(6.63)	\\
6151.620	&		&	2.180	&	-3.299	&	45.7(6.46)	\\
6180.200	&		&	2.730	&	-2.780	&	43.8(6.56)	\\
6232.640	&		&	3.650	&	-1.270	&	50.1(6.29)	\\
6240.650	&		&	2.220	&	-3.170	&	54.3(6.58)	\\
6254.250	&		&	2.280	&	-2.400	&	71.6(6.30)	\\
6419.940	&		&	4.730	&	-0.240	&	50.4(6.53)	\\
6481.870	&		&	2.280	&	-2.984	&	57.0(6.50)	\\
6575.020	&		&	2.590	&	-2.820	&	44.6(6.42)	\\
6750.150	&		&	2.420	&	-2.620	&	60.8(6.37)	\\
4520.220	&	Fe II&	2.810	&	-2.600	&	54.4(6.45)	\\
6247.557	&		&	3.892	&	-2.329	&	24.5(6.45)	\\
\hline
\end{tabular}}

\small The number in the parenthesis gives the derived abundance from the respective line. 
\end{table*}

\begin{table*}
\centering
{\bf Table A2: Equivalent widths (in m\r{A}) of lines used for deriving elemental abundances}
\resizebox{0.68\textwidth}{!}{\begin{tabular}{ccccc}
\hline                       
Wavelength(\r{A}) & Element & $E_{low}$(eV) & log gf & LAMOST J045019.27+394758.7 \\
\hline 
5682.630	&	Na I	&	2.100	&	-0.700	&	89.6(5.99)	\\
6154.230	&		&	2.100	&	-1.560	&	37.0(5.84)	\\
6160.750	&		&	2.100	&	-1.260	&	49.3(5.78)	\\
4730.030	&	Mg I	&	4.350	&	-2.523	&	31.6(6.77)	\\
5711.090	&		&	4.350	&	-1.833	&	54.1(6.49)	\\
5690.430	&	Si I	&	4.930	&	-1.870	&	25.9(6.68)	\\
5772.150	&		&	5.080	&	-1.750	&	23.6(6.67)	\\
5948.540	&		&	5.080	&	-1.230	&	42.8(6.59)	\\
6131.850	&		&	5.620	&	-1.140	&	23.4(6.64)	\\
6142.490	&		&	5.620	&	-1.480	&	19.1(6.87)	\\
6155.140	&		&	5.620	&	-0.770	&	31.7(6.48)	\\
6243.820	&		&	5.610	&	-1.100	&	21.5(6.55)	\\
5261.710	&	Ca I	&	2.520	&	-0.730	&	55.1(5.26)	\\
5581.970	&		&	2.520	&	-0.710	&	57.5(5.28)	\\
5590.110	&		&	2.520	&	-0.710	&	58.8(5.31)	\\
6166.440	&		&	2.520	&	-0.900	&	53.0(5.33)	\\
6455.600	&		&	2.520	&	-1.350	&	41.7(5.50)	\\
6471.660	&		&	2.530	&	-0.590	&	65.5(5.30)	\\
6499.650	&		&	2.520	&	-0.590	&	62.6(5.23)	\\
4453.710	&	Ti I	&	1.870	&	-0.010	&	26.7(3.73)	\\
4759.270	&		&	2.250	&	0.514	&	32.9(3.78)	\\
4840.870	&		&	0.900	&	-0.509	&	46.9(3.55)	\\
5210.390	&		&	0.050	&	-0.884	&	72.0(3.53)	\\
4568.310	&	Ti II	&	1.220	&	-2.650	&	25.6(3.57)	\\
4764.530	&		&	1.240	&	-2.770	&	31.2(3.84)	\\
4798.520	&		&	1.080	&	-2.430	&	46.6(3.72)	\\
4865.610	&		&	1.120	&	-2.610	&	29.9(3.50)	\\
5381.020	&		&	1.570	&	-2.080	&	34.9(3.59)	\\
5737.060	&	V I	&	1.060	&	-0.740	&	41.6(3.76)	\\
6039.720	&		&	1.060	&	-0.650	&	33.5(3.45)	\\
6292.820	&		&	0.290	&	-1.470	&	48.4(3.66)	\\
6531.420	&		&	1.220	&	-0.840	&	30.5(3.72)	\\
4737.380	&	Cr I	&	3.090	&	-0.099	&	33.0(4.55)	\\
4870.800	&		&	3.080	&	0.050	&	34.6(4.43)	\\
4942.490	&		&	0.940	&	-2.294	&	43.8(4.50)	\\
5247.570	&		&	0.960	&	-1.640	&	58.7(4.25)	\\
5298.280	&		&	0.980	&	-1.150	&	79.3(4.38)	\\
5300.740	&		&	0.980	&	-2.120	&	56.0(4.67)	\\
4634.070	&	Cr II	&	4.070	&	-1.240	&	38.8(4.88)	\\
4812.350	&		&	3.860	&	-1.140	&	40.1(4.59)	\\
4703.800	&	Ni I	&	3.660	&	-0.735	&	33.6(5.16)	\\
4732.460	&		&	4.110	&	-0.550	&	32.3(5.44)	\\
4752.410	&		&	3.660	&	-0.700	&	45.0(5.44)	\\
4852.560	&		&	3.540	&	-1.070	&	26.8(5.17)	\\
4937.340	&		&	3.610	&	-0.390	&	49.2(5.18)	\\
4953.200	&		&	3.740	&	-0.670	&	34.5(5.19)	\\
6086.280	&		&	4.270	&	-0.530	&	28.5(5.45)	\\
6111.070	&		&	4.090	&	-0.785	&	28.0(5.48)	\\
6175.360	&		&	4.090	&	-0.530	&	33.9(5.39)	\\
6186.710	&		&	4.110	&	-0.777	&	20.8(5.29)	\\
6327.590	&		&	1.680	&	-3.150	&	41.3(5.33)	\\
6643.640	&		&	1.680	&	-2.300	&	69.0(5.19)	\\
\hline
\end{tabular}}

The number in the parenthesis gives the derived abundance from the respective line. 

\end{table*}

\begin{table*}
\centering
\resizebox{0.68\textwidth}{!}{\begin{tabular}{ccccc}
\hline                       
Wavelength(\r{A}) & Element & $E_{low}$(eV) & log gf & LAMOST J045019.27+394758.7 \\
\hline
4722.150	&	Zn I	&	4.030	&	-0.370	&	26.9(2.97)	\\
4810.530	&		&	4.080	&	-0.150	&	27.3(2.81)	\\
5119.110	&	Y II	&	0.990	&	-1.360	&	13.3(1.06)	\\
5200.410	&		&	0.990	&	-0.570	&	37.7(1.07)	\\
5402.770	&		&	1.840	&	-0.510	&	12.9(1.15)	\\
4486.910	&	Ce II	&	0.290	&	-0.180	&	22.6(0.51)	\\
4562.360	&		&	0.480	&	0.210	&	19.9(0.23)	\\
4628.160	&		&	0.520	&	0.008	&	18.6(0.43)	\\
5292.620	&	Pr II	&	0.650	&	-1.005	&	28.1(1.73)	\\
6165.890	&		&	0.920	&	-0.205	&	18.9(0.90)	\\
4451.560	&	Nd II	&	0.380	&	0.070	&	39.0(0.76)	\\
4797.150	&		&	0.560	&	-0.690	&	19.4(0.98)	\\
4811.340	&		&	0.060	&	-1.140	&	23.4(0.98)	\\
4859.040	&		&	0.320	&	-0.440	&	32.8(0.90)	\\
5130.590	&		&	1.300	&	0.450	&	29.3(1.03)	\\
4434.320	&	Sm II	&	0.380	&	-0.070	&	14.8(-0.26)	\\
4566.200	&		&	0.330	&	-0.590	&	11.4(0.04)	\\
4674.590	&		&	0.180	&	-0.560	&	22.7(0.27)	\\
4676.900	&		&	0.040	&	-0.870	&	11.7(-0.02)	\\
\hline
\end{tabular}}

The number in the parenthesis gives the derived abundance from the respective line. 
\end{table*}

\end{document}